\input harvmac
\input mssymb.tex
\input labeldefs.tmp
\writedefs
\overfullrule=0pt

\input epsf

\def\fig#1#2#3{
\xdef#1{\the\figno}
\writedef{#1\leftbracket \the\figno}
\nobreak
\par\begingroup\parindent=0pt\leftskip=1cm\rightskip=1cm\parindent=0pt
\baselineskip=11pt
\midinsert
\centerline{\epsfbox{#3}}
\vskip 12pt
{\bf Fig. \the\figno:} #2\par
\endinsert\endgroup\par
\goodbreak
\global\advance\figno by1
}

\def\rhot{\tilde{\rho}}
\def\Im{\mathop{\rm Im}\nolimits}
\def\blob{\indent $\bullet$ }
\def\Egs{E_{\rm g.s.}}
\def\Ghat{\widehat{G}}
\def\Chat{\widehat{C}}

\font\email=cmtt9

\Title{\vbox{\baselineskip12pt\hbox{LPTENS 97/39}\hbox{cond-mat/9801158}}}
{\vbox{\centerline{The generalized multi-channel Kondo model:}
\vskip2pt\centerline{Thermodynamics and fusion equations}}}
\centerline{P.~Zinn-Justin\footnote{$^{(1)}$}{{\email pzinn@physique.ens.fr}}
and
N.~Andrei\footnote{$^{(2)}$}{{\email natan@physics.rutgers.edu}}\footnote{$^\dagger$}{Permanent address:
Department of Physics and Astronomy, Rutgers University,
Piscataway, NJ 08855, USA}
}
\medskip\centerline{Laboratoire de Physique Th\'eorique de l'Ecole
Normale Sup\'erieure\footnote{*}{unit\'e
propre du CNRS, associ\'ee \`a l'Ecole Normale
Sup\'erieure et l'Universit\'e Paris-Sud.}}
\centerline{24 rue Lhomond, 75231
Paris Cedex 05, France}

\vskip .3in
\centerline{Abstract}
The $SU(N)$ generalization of the multi-channel Kondo model
with arbitrary rectangular impurity representations is
considered by means of the Bethe Ansatz.
The thermodynamics of the model is analyzed by
introducing modified fusion equations for the impurity, leading to
a simple description of
the different IR fixed points of the theory. The entropy
at zero temperature is discussed; in particular the
overscreened case is explained in terms of quantum group representation.

\Date{1/98}

\newsec{Introduction and basic equations}
The Kondo model with antiferromagnetic coupling was studied 
in 1964~\ref\KON{J.~Kondo, {\it
Progr. Theor. Phys.} 32 (1964), 37.} and found account for the low-temperature
resistivity of metals in which
magnetic impurities were present (Kondo effect). The perturbative 
calculations led to divergences as the temperature was lowered,
indicating an inaccessible strong-coupling IR region. This
stimulated great theoretical interest in the
field. Several approaches were used to study the Kondo effect
in more detail\nref\AND{P.W.~Anderson, {\it J. Phys.} C3 (1970),
2346.}\nref\NOZ{P.~Nozi\`eres, {\it Proc. of 14th Int. Conf.
on Low Temp. Physics} (1975), Ed. M.~Krusius and M.~Vuorio.}\nref
\WIL{K.G.~Wilson, {\it Rev. Mod. Phys.} 47 (1975), 773.}
[\xref\AND--\xref\WIL];
here we are particularly interested in the fact that the model,  
formulated in terms of one dimensional fields, was found
to be solvable by Bethe Ansatz techniques\nref
\NAT{N.~Andrei, {\it Phys. Rev. Lett.} 45 (1980), 379.
}\nref\WI{P.~Wiegmann, {\it Sov. Phys. J.E.T.P. Lett.} 31 (1980),
392.}
[\xref\NAT,\xref\WI].

In this paper we study the solution of a
generalized Kondo model, describing fermions carrying color (spin) and flavor
(channel) in
the fundamental representations of $SU(N)$ and $SU(f)$ respectively, interacting
with a localized impurity. The impurity
is in a $(n \times l)$ rectangular representation of the color,
and in a trivial representation of the flavor group. The model combines and
generalizes the
multichannel Kondo Model of Nozieres and
Blandin~\ref\NOZB{P.~Nozi\`eres and A.~Blandin, {\it J. Phys. (Paris)}
41 (1980), 193.} applicable when the impurity
orbital structure is taken into account, and the $SU(N)$ version of Coqblin and
Schrieffer~\ref\COQS{B. Coqblin and J. ~Schrieffer, {\it Phys. Rev.}
185 (1969),847.} valid for rare earth materials with strong $j-j$ coupling.

We shall diagonalize the model (section 1); the Bethe-Ansatz basis separates 
in a natural way into charge, spin, and flavor sectors.  This separation,
when the interaction with the impurity is turned off,
corresponds to the decomposition of the free fermion
CFT describing the electrons into
a $U(1)$ sector, a $SU(N)$ level $f$ WZW and a $SU(f)$ level $N$ WZW.
Only the spin sector interacts non-trivially
with the impurity as we shall see by studying the associated Bethe
Ansatz Equations.
In this sector, we shall find that two symmetries play a key role:
the original $SU(N)$ symmetry of the whole theory, and a new hidden
quantum group $SU_q(N)$ symmetry where $q=\exp(2i\pi/(f+N))$.
These symmetries will naturally appear
in the structure of the effective Bethe Ansatz Equations describing
this sector. It should be noted that we are here extending
the well-known connection
between the WZW $SU(N)$ level $f$ theory, i.e.
the non-interacting spin sector of the electrons,
with quantum group $SU_q(N)$ -- through the braiding matrices
or the truncated fusion rules -- to our interacting (non conformal)
theory: the fact that the
quantum group symmetry is preserved is probably equivalent to the fact
that integrability is preserved.

We shall next briefly study the ground state and the low-lying
excitations (section 2), and then move on to
the thermodynamics of the system. In order to do so we shall
consider a scaling limit procedure where the linear density of the
electrons $D$, which can be thought of the order of the depth
of the Fermi sea, plays the role of a UV cutoff of the theory and is
sent to infinity while keeping the physical energy scale,
$De^{-2\pi/Nc}$, finite. We can then introduce {\it fusion
equations} dressed by finite temperature for the impurity; we quote
equations \fus:
\eqn\ifus{\chi^r_j(\zeta+i\pi/N) \chi^r_j(\zeta-i\pi/N)
=\chi^{r+1}_j(\zeta) \chi^{r-1}_j(\zeta) 
+ \chi^r_{j+1}(\zeta) \chi^r_{j-1}(\zeta) 
e^{-2\delta_{jf} \sin(\pi r/N) e^\zeta}.}
Here $\zeta$ is a parametrization of the temperature
($T\propto e^{-\zeta}$) and
$\chi^r_j$ is the contribution of the impurity whose representation
is a rectangular $r\times j$ Young tableau
to the (finite temperature) partition function. So this system of
equations connects impurities with differents spins (intuitively
impurities with high spin are made out of smaller spin impurities
through the fusion procedure), and it contains
all the physics of the model. Indeed we shall use it to compute
physically interesting quantities in the low and high temperature
regimes (section 4). Finally we shall tackle with the interpretation of the
zero temperature entropy (section 5) using quantum group arguments.

\subsec{The model.}
\def\frac#1#2{{#1\over #2}}

The Hamiltonian to be studied is,

\eqn\hami{
{\cal H}=-i\int \bar{\psi}^a_i(x) \partial_x \psi^a_i(x) dx+J
\bar{\psi}^a_i(0) \sigma_A^{ab} \psi^b_i(0) \bar{\chi}^{\alpha}\tau_A^{
\alpha \beta}\chi^{\beta}. 
}

The energy spectrum of the electrons is taken to be linear in the 
momentum as we shall be
studying universal properties  near the Fermi surface. 

The electrons are in the fundamental representation  $\sigma_A$ 
(corresponding to Young
tableau consisting of one box), and the impurity in the representation
$\tau_A$, given by a rectangular Young tableau
with $l$ columns and $n$ rows, $1\le n\le N-1$.
The impurity is a singlet of the flavor group $SU(f)$, while the
electrons are in its fundamental representation.
The fermionic field $\psi^{a}_{i}(x)$ annihilates an electron
at $x$ with spin (or color) index
$a$, $a=1,...,N$ and flavor index $i$, $i=1,...,f$. Note that from the
point of view of 2D quantum field theory, there is only one chirality
of electrons (``right-movers''). The operator 
$\bar{\chi}^{\alpha}\tau_A^{
\alpha \beta}\chi^{\beta}$ represents the impurity spin 
operator in a
representation $\{\tau_A\}$, where the impurity field $\chi^\alpha$
is taken to be fermionic and subject 
to the constraint $ \sum_\alpha \bar{\chi}^\alpha\chi^\alpha=1$.
Summation over all indices is implied, $A=1\ldots N^2-1$
being a $\goth{su}(N)$ Lie algebra index.

The interaction with the impurity breaks the $U(fN)$ symmetry of the free
hamiltonian down to
$U(1)^{charge}\times SU(N)^{spin} \times SU(f)^{flavor}$. It will be
implicitly assumed that all the flavor levels are equally populated.

We shall find that the model possesses a variety of IR fixed points,
whose nature depends on 
the symmetry structure in the flavor sector and on the spin representation 
$n \times l$, generalizing
the familiar  $N=2$ case (the multichannel Kondo model
\nref\NOZB{P.~Nozi\`eres and A.~Blandin, {\it J. Phys. (Paris)}
41 (1980), 193.}\nref\AD{N.~Andrei and C.~Destri, {\it Phys. Rev. Lett.}
52 (1984), 364.}\nref\TW{A.~Tsvelik and P.~Wiegmann, {\it J. Phys.} 54 (1984), 201.}
[\xref\NOZB--\xref\TW]).
We shall identify the mechanism underlying
the appearance of these
fixed points as {\it dynamical fusion} by which the electrons form 
spin complexes whose
interaction with the impurity leads to a new behavior in the infrared
\AD. These complexes consist of $f$ electrons fused 
into local objects that transform
according to Young Tableaux of one row of length $f$. These composites
interact with the impurity
and determine low energy properties of the model.

In the Bethe-Ansatz approach a precise description of the formation
of these composites can be given. To do so, a careful cut-off
procedure needs to be introduced to allow the formation
of the composites while maintaining integrability in the presence of
a finite cut-off. The  linearized hamiltonian propagates separately
the charge-spin-flavor degrees of freedom  that make up the electron.
Therefore the effect of flavor on the spin degrees of freedom is is 
recovered only in the full space. To follow the dynamic 
coupling of spin and flavor we add some curvature which maintains the identity
of the electron while allowing its components to interact. In the end of
the calculation
the cut-off is sent to infinity and the curvature removed. Already in the free 
field theory the resulting theory is quite involved, and even the counting of
states is not trivial~\ref\DES{C.~Destri, Unpublished Notes (1985).}. 
However, the resulting basis is the natural one in which
to turn on the impurity, it is the zero order approximation in the sense of
degenerate perturbation theory.

The scheme consists of the following elements:

\bigskip
\item{$\diamond$}
A second derivative term with a cutoff, $\Lambda$,
\eqn\ct{
{\cal H}_{\Lambda} = -\frac{1}{2 \Lambda} 
\int \bar{\psi}_{i}^a (x) \partial_x^2
\psi_i^a (x)dx.
}
This term explicitly provides an energy cut-off.
Furthermore, it introduces  curvature into the electronic spectrum 
and breaks charge-spin-flavor (CSF) separation.
Once the electron composites
are formed, and the low-energy spectrum of the theory is identified,
the cutoff is taken to infinity.\hfil\break 
Adding the term \ct\ also imposes
restrictions on the form of the eigenstates which can be expressed 
in terms of the following counterterms without which 
the model is not integrable for finite $\Lambda$.

\item{$\diamond$}
An electron-electron interaction term, of the form
\eqn\jt{
2\tilde{J}
\int 
\bar{\psi}_i^a (x) \bar{\psi}_{j}^{b}(x) \psi_{i}^{b} (x)
\psi_{j}^{a} (x)dx , 
}
The term
has no effect on the spectrum once the cut off is removed
and no impurity is present, independently of the value of
the coupling $\tilde{J}$. The
linearized spectrum has a large degeneracy which
is removed when the interaction with the impurity is added. The
addition of \ct\ and \jt\ provides a way of finding the
eigenstates, as we will show below.

\item{$\diamond$}
A counterterm ${\cal H}_{cc}$, of the form 
\eqn\hcc{
{\cal H}_{cc} = \frac{1}{\Lambda}
\int \bar{\psi}_{i}^a(x) V(x)\psi_{i}^a (x) dx, 
}
with
\eqn\vx{
V(x)=\frac x{|x|}(\delta ^{\prime }(x^{+0})+\delta ^{\prime }(x^{-0})),
}
needs to be added to the Hamiltonian in order to preserve
integrability at the origin; this term vanishes 
once the cutoff is removed, and plays no further role in the problem.

Eigenstates of \hami\ with $N_e$ electrons and one impurity are
of the form
\eqn\eig{
|F> =  \int
\prod_{j} dx_j\, F_{\{i_j\},b}^{\{a_j\}}(\{x_j\})
\bar{\chi}_b 
\prod_{j=1}^{N_e} \bar{\psi}^{a_j}_{i_j}(x_j) |0>,
}
where the fermionic field $\bar{\chi}_b$ creates the impurity at $x=0$.
The amplitude $F$
satifies the differential equation $h|F>=E|F>$, where the first quantized
Hamiltonian $h$ takes the form 
\eqn\eigb{
\eqalign{
h &= \sum_{j=1}^{N^e} \{-i \partial_j - \frac{1}{2\Lambda} \partial_j^2 
+ 2J\delta(x_j) \sigma_A \tau_A \}
\cr  &+ \sum_{l<j} 2 \tilde{J} \delta(x_l-x_j)
(P_{lj}-{\cal P}_{jl}) + \sum_{j=1}^{N^e} \frac{1}{\Lambda} V(x_j),
}}
with $P_{jl}$(${\cal P}_{jl}$) being the spin (flavor) exchange operator.
When the impurity is also in the fundamental representation, we can
write
\eqn\eigc{
\eqalign{
h &= \sum_{j=1}^{N^e} (-i \partial_j - (\Lambda^{-1}) \partial_j^2 
+ 2J \delta(x_j) P_{j0}) 
\cr &+ \sum_{l<j} 2 \tilde{J} \delta(x_l-x_j)
(P_{lj}-{\cal P}_{jl}) +  \sum_{j=1}^{N^e} \frac{1}{\Lambda} V(x_j).
}}

We see that the interaction terms act only when electrons coincide at
the same point or at the impurity site. Hence, the eigenstate
amplitudes are combinations of plane waves with pseudo-momenta $k_j$,
and have coefficients that depend on the ordering of the electrons and
the spin and flavor indices. These coefficients are related through
products of electron-electron and electron-impurity $S$-matrices that we
will write below. Here we will only write explicitely the results for
an impurity in the fundamental representation.

The electron-impurity $S$-matrix can be written, to first order in
$1/\Lambda$, 
\eqn\eis{
S_{j0} = e^{i \arctan \frac{c}{1+\lambda_j}}
\left(\frac{\lambda_j+1-ic P_{j0}}{\lambda_j+1-ic}\right), 
}
where
\eqn\eisb{
\lambda_j = \left( \frac{1+J^2}{1-J^2} \right) \frac{k_j}{\Lambda},
\quad c
\equiv \frac{2J}{1-J^2}.
}
In the scaling limit, $J$ and $c$ have the same scaling
behavior. Notice that \eis\ is trivial in the flavor sector. 

The electron-electron $S$-matrix is of the form
\eqn\ees{
S_{jl} = \frac{\lambda_j-\lambda_l-ic
P_{jl}}{\lambda_j-\lambda_l-ic}  \frac{\lambda_j-\lambda_l+ic
{\cal P}_{jl}}{\lambda_j-\lambda_l+ic}.
}
if we set
\eqn\tJ{
\tilde{J} = \frac{J}{1+J^2},
}
Integrability is guaranteed since $S_{j0}$ and $S_{jl}$ satisfy the
Yang-Baxter conditions
\eqn\YBE{
\eqalign{
S_{jl} S_{j0} S_{l0} &= S_{l0} S_{j0} S_{jl} \cr
S_{jl} S_{jk} S_{lk} &= S_{lk} S_{jk} S_{jl}
}}

Finally, the energy eigenvalue of a $N_e$ electron state is of the form
\eqn\oen{
E = \sum_{j=1}^{N^e} k_j(1+\frac{k_j}{2\Lambda}).
}

In order to determine the spectrum, we impose periodic
boundary conditions, and solve the corresponding eigenvalue problem. 
The procedure is standard~\ref\LEC{N.~Andrei,
{\it Integrable Models in Condensed Matter Physics},
in {\it Series on Modern Condensed Matter Physics - Vol. 6}, 458-551,
World Scientific, Lecture Notes of ICTP Summer Course (September
1992). Editors: S. Lundquist, G. Morandi, and Yu Lu.}
and we skip here the details.
The result is contained in
the Bethe Ansatz Equations (B.A.E.) which we proceed to write down. 
Each of the degrees of freedom -- charge, spin and flavor --
is described by a set of variables whose number depends on the symmetry
of the particular state.
The charge degrees of freedom are given by the set
$\{k_j,\, j=1,...,N^e\}$. The spin degrees of
freedom are parametrized by the sets $\{\Lambda _\gamma ^r,\, \gamma
=1,\ldots,M^r,\, r=1,\ldots,N-1\}$. 
Finally, the flavor degrees of freedom
are represented by the sets $\{\omega _\gamma ^r,\, \gamma =1,\ldots,\bar{M}
^r,\, r=1,\ldots,f-1\}$. The set of integers $\{ M^r,\,
r=1,\ldots,N-1\}$
specify the symmetry of the spin component of the wave function.
Similarly, the quantum numbers
$\{\bar{M}^r\}$ specify the symmetry of the  flavor component.

The equations are: ($L$ size of the periodic space)
\eqn\oBAE{
\eqalign{
&\qquad e^{ik_jL} =\prod_{\gamma =1}^{M^1}\frac{\Lambda _\gamma ^1-(1+\lambda
_j)+i\frac c2}{\Lambda _\gamma ^1-(1+\lambda _j)-i\frac c2}\prod_{\gamma =1}^{
\bar{M}^1}\frac{\omega _\gamma ^1-(1+\lambda _j)-i\frac c2}{\omega _\gamma
^1-(1+\lambda _j)+i\frac c2}, \cr
&\left\lbrace\eqalign{
-\prod_{\beta =1}^{\bar{M}^1}\frac{\omega _\gamma ^1-\omega _\beta ^1+ic}{%
\omega _\gamma ^1-\omega _\beta ^1-ic} &= \prod_{j=1}^{N^e}\frac{\omega
_\gamma ^1-(1+\lambda_j)+i\frac c2}{\omega _\gamma ^1-(1+\lambda _j)-i\frac c2}
 \prod_{\beta =1}^{\bar{M}^2}\frac{\omega _\gamma
^1-\omega _\beta ^2+i\frac 
c2}{\omega _\gamma ^1-\omega _\beta ^2-i\frac c2},\cr
-\prod_{\beta =1}^{\bar{M}^r}\frac{\omega _\gamma ^r-\omega _\beta ^r+ic}{%
\omega _\gamma ^r-\omega _\beta ^r-ic} &=\prod_{t=r\pm 1}\prod_{\beta =1}^{%
\bar{M}^t}\frac{\omega _\gamma ^r-\omega _\beta ^t+i\frac c2}{\omega _\gamma
^r-\omega _\beta ^t-i\frac c2} \qquad r=2,\ldots,f-1, \cr}\right.\cr
&\left\lbrace\eqalign{
-\prod_{\beta =1}^{M^1}\frac{\Lambda _\gamma ^1-\Lambda _\beta ^1+ic}
{\Lambda_\gamma^1-\Lambda _\beta ^1-ic}
&= \prod_{j=1}^{N^e}\frac{\Lambda _\gamma ^1-(1+\lambda _j)+i\frac c2}
{\Lambda _\gamma ^1-(1+\lambda _j)-i\frac c2}
\prod_{\beta =1}^{M^2}\frac{\Lambda_\gamma ^1-\Lambda _\beta ^2+i\frac c2}
{\Lambda _\gamma ^1-\Lambda_\beta ^2-i\frac c2},\cr
-\prod_{\beta =1}^{M^n}\frac{\Lambda _\gamma ^n-\Lambda _\beta ^n+ic}{\Lambda _\gamma
^n-\Lambda _\beta ^n-ic} &= 
\frac{\Lambda _\gamma ^n+il\frac c2}{\Lambda _\gamma
^n-il\frac c2}
\prod_{t=n\pm 1}\prod_{\beta =1}^{M^t}\frac{\Lambda
_\gamma ^n-\Lambda _\beta ^t+i\frac c2}{\Lambda _\gamma ^n-\Lambda _\beta ^t-i
\frac c2},\cr
-\prod_{\beta =1}^{M^r}\frac{\Lambda _\gamma ^r-\Lambda _\beta ^r+ic}{\Lambda _\gamma
^r-\Lambda _\beta ^r-ic} &= \prod_{t=r\pm 1}\prod_{\beta =1}^{M^t}\frac{\Lambda
_\gamma ^r-\Lambda _\beta ^t+i\frac c2}{\Lambda _\gamma ^r-\Lambda _\beta ^t-i
\frac c2} \qquad r=2,\ldots,N-1, r\ne n. \cr
}\right.
}
}

The next step is to solve the equations for all possible states,
identify the ground state and the low energy
excitations above it. Subsequently, by summing over all excitation energies
we shall obtain the partition function.

The B.A.E. are a function of the cutoff $\Lambda$ which
eventually is sent to infinity. In this
limit  the equations reduce to a smaller set once the correct ground
state has been identified. It corresponds to {\it string} solutions
(see below) leading to electron composites
which interact most efficiently with the impurity.
The ground state and low lying excitations lie
in a sector of the theory given by solutions of a particular form --
$f$-$strings$. Solutions of this type are $SU(f)$ flavor singlets
which allows them to have maximally large $SU(N)$ spin. We shall see that
this class of 
excitations is characterized by a scale 
$T_0=De^{-\frac{2\pi }{N c}}$. When strings are broken to form
flavored excitations we expect them to be characterized by other
scales which will tend to infinity as the cut-off is removed and
thus do not contribute to the impurity dynamics.

The formation of composites in flavor corresponds to solutions of the
B.A.E. where
the charge parameters, 
$\{\lambda_j\}$, are complex numbers centered around $\{\omega
_\gamma ^1\}$. Likewise, rank $r$ 
flavor parameters are themselves centered around rank $r+1$ solutions
\AD. The form of the charge parameters is, 
\eqn\flstr{
\{\lambda_j,\, j=1\ldots N_e\}
=\left\{
p_\delta/\Lambda +ic\left( \frac{f +1}2-q\right)
;\ q=1,2,\ldots,f ,\ p_\delta\ \hbox{\rm real},\ \delta=1,...,N^e/f\right\}. 
}
while the flavor parameters,
\eqn\flstrb{
\{\omega^r_\gamma,\, \gamma=1,2,..., M^r\}
=\left\{p_\delta/\Lambda + iJ\left( {f-r+1\over 2} -q\right);\ 
q=1,2,\ldots,f-r, \ \delta=1,\ldots,N^e/f\right\}
}
where $r=1,...,f-1$. These configurations satisfy the B.A.E.
in a trivial manner
and induce {\it fusion} in the B.A.E. equations as well as in the form of the
wavefunctions. A string built on momentum $p$ as its real part induces in the 
wave function a composite of the form $exp\{-\frac{1}{2}\Lambda J  \sum_{j,l} |x_j-x_l| +
ip(x_1+...  +x_f)\} \times [\ldots] $, which becomes local as $\Lambda \rightarrow \infty$.
These composites will be described by effective {\it fused} B.A. Equations obtained
by inserting the string configurations into the full B.A.E.  After removing the cutoff 
they become
\eqn\BAE{\eqalign{
\prod_{\scriptstyle\beta=1\atop\scriptstyle\beta\ne\gamma}^{M^r} 
{\Lambda_\gamma^r-\Lambda_\beta^r+ic
\over\Lambda_\gamma^r-\Lambda_\beta^r-ic}
\,
\prod_{t=r\pm 1} &\prod_{\beta=1}^{M^t}
{\Lambda_\gamma^r-\Lambda_\beta^t-ic/2
\over\Lambda_\gamma^r-\Lambda_\beta^t+ic/2}\cr
&=
\left({\Lambda_\gamma^1-1+ifc/2\over\Lambda_\gamma^1-1-ifc/2}\right)^{N_e\delta^{r1}}
\left({\Lambda_\gamma^n+ilc/2\over\Lambda_\gamma^n-ilc/2}\right)^{\delta^{rn}}\cr
}}
for each root $\Lambda^r_\gamma$.
$N_e$ is now the number of composites of $f$ electrons
(i.e. originally $N_e/f$). \BAE\ will be our starting point for all
subsequent calculations.

The energy $\varepsilon$ of a composite of $f$ electrons is given by:
\eqn\en{e^{iL\varepsilon}=\prod_{\gamma=1}^{M^1}
\left({\Lambda^1_\gamma-1+ifc/2 \over \Lambda^1_\gamma-1-ifc/2}\right)}

The (irreducible)
representation $R$ of the resulting state is given by the numbers
$M^r$ of roots of type $r$;
indeed, each root
$\Lambda_\gamma^r$ moves one box of the Young tableau down from row $r$
to row $r+1$. In other words one starts with the empty configuration
$M^r=0$ which corresponds to the highest possible weight (fig.~\hiw).
\fig\hiw{The highest possible weight of the model.}{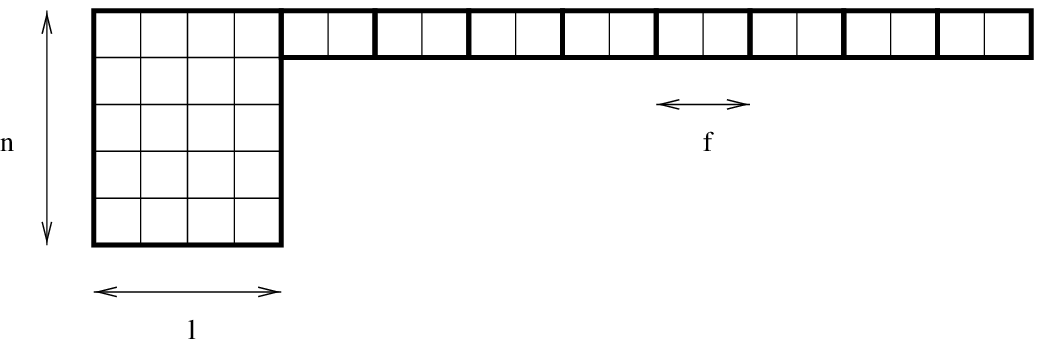} 
We call this representation $R_0$.
Then
\eqn\rep{R=R_0-\sum_{r=1}^{N-1} M^r \alpha^r}
where $\alpha^r=e^r-e^{r+1}$ is the $r^{\rm th}$ simple root of
the Dynkin diagram $A_{N-1}$ 
($e^a$, $a=1\ldots N$ corresponds to one box in row $a$ of
the corresponding Young diagram,
see appendix A for more details).

\subsec{Continuous B.A.E.} As we are mainly interested in the
thermodynamics of this model, we shall immediately write equations
in the thermodynamic limit
($N_e\rightarrow\infty$, keeping the density of electrons
per unit length $D\equiv N_e/L$ and 
the densities of roots $M^r/L$ fixed i.e. of the order
of some physical energy scale). In this limit, standard calculations lead
to continuous B.A.E. Let us briefly
derive these in a rather formal way which will minimize the amount
of calculations.

We recall that according to the ``string hypothesis'',
the $\Lambda_\gamma^r$ group into strings
of roots with the same real part, and a fixed distance of $ic$ between
two consecutive roots.

We therefore introduce some notations. We shall need the function 
\eqn\defth{\Theta(\Lambda)\equiv {1\over
i}\log{\Lambda-ic/2\over\Lambda+ic/2}=
2 \arctan\left({2\Lambda\over c}\right) + \pi}
where we have specified the determination of the log on the real axis,
so that $\Theta(-\infty)=0$.

We also define its ``descendants'' generated by strings:
\eqn\defthd{\Theta_{k_1,\ldots,k_p}(\Lambda)=
\sum_{j_1,\ldots,j_p} \Theta(\Lambda+ic(j_1+\cdots+j_p)),}
each $j_a$ being an integer (or a half-integer, depending on the
parity of $k_a$) which ranges from $-(k_a-1)/2$ to $(k_a-1)/2$.
In particular $\Theta_k(\Lambda)=(1/i)\log((\Lambda-ikc/2)/(\Lambda+ikc/2))$.
Again, on the real axis, the analytic structure is chosen so that no
cut crosses the real axis and $\Theta_{k_1,\ldots,k_p}(-\infty)=0$.

In the same way we define 
\eqn\defK{K(\Lambda)= {1\over 2\pi} {d\over d\Lambda} \Theta(\Lambda)={1\over 2\pi}{c\over\Lambda^2+c^2/4}}
and its descendants $K_{k_1,\ldots,k_p}$.

If $\Lambda^r_{j;\gamma}$ is the center of
the $\gamma^{\rm th}$ string ($1\le \gamma\le M^r_j$) 
of length $j$ ($j$ positive integer)
and of type $r$ ($1\le r\le N-1$), then by multiplying eq. \BAE\ for
all the $j$ roots of a given string and taking the log, we find:
\eqn\logBAE{\eqalign{
2\pi I^r_{j;\gamma}&=\delta^{r1} N_e
\Theta_{f,j}(\Lambda^r_{j;\gamma}-1)
+\delta^{rn} \Theta_{l,j}(\Lambda^r_{j;\gamma})\cr
&-\sum_{k=1}^\infty
\sum_{\beta=1}^{M^r_k}\Theta_{2,j,k}(\Lambda^r_{j;\gamma}
-\Lambda^r_{k;\beta})
+\sum_{t=r\pm 1}\sum_{k=1}^\infty\sum_{\beta=1}^{M^t_k}
\Theta_{j,k}(\Lambda^r_{j;\gamma}-\Lambda^t_{k;\beta})\cr}.}
The $I^r_{j;\gamma}$ are half-integers of a given parity.
Differentiating once with respect to $\Lambda$ and introducing the
densities of roots and holes
$\rho^r_j$ and $\rhot^r_j$ we get:
\eqn\coBAEa{\rhot^r_j+\rho^r_j=\delta^{r1} N_e K_{f,j}(\Lambda-1)
+\delta^{rn} K_{l,j}(\Lambda)
-\sum_k K_{2,j,k}\star\rho^r_j+\sum_{t=r\pm 1}\sum_k K_{j,k}\star\rho^t_k}
where $\star$ denotes convolution in $\Lambda$ space. From now on,
unless explicitly stated otherwise, all
subscript indices are string indices, running from $1$ to $\infty$,
and all superscript indices are Dynkin diagram indices, running from
$1$ to $N-1$.

$\rho^r_j$ is normalized by the condition 
$\sum_j j\int d\Lambda\, \rho^r_j=\sum_j j\, M^r_j
=M^r$. This is slightly incorrect since the $M^r$ diverge in the
thermodynamic limit, only $M^r/L$ is physical; but it simplifies notations.

To simplify even further the form of Eq. \coBAEa\ let us introduce the
``spectral parameter-dependent'' Cartan matrices $C^{qr}$, $C_{jk}$ 
of $A_{N-1}$, $A_\infty$ and their inverses $G^{qr}$,
$G_{jk}$.

For $1\le q,r\le N-1$, $C^{qr}$ is defined by
\eqn\defC{C^{qr}(\Lambda)=\delta^{qr}\delta(\Lambda)-(\delta^{qr+1}
+\delta^{qr-1})s(\Lambda)}
where $s$ is the simple function $s(\Lambda)=1/(2c\,\cosh(\pi
\Lambda/c))$.
The same formula holds for $C_{jk}$, only the boundary conditions
$1\le j,k<\infty$ being different.

In order to express $G^{qr}$ and $G_{jk}$ we use Fourier transform
defined by
$$\phi(\kappa)\equiv \int \phi(\Lambda)
\exp(i\kappa\, 2\Lambda/c) d\Lambda$$
for any function $\phi(\Lambda)$. We have
$s(\kappa)=1/(2\cosh(\kappa))$, and
\eqn\defG{\eqalign{
G^{qr}(\kappa)&=G^{rq}(\kappa)=2 \coth(\kappa) 
{\sinh((N-q)\kappa) \sinh(r\kappa)
\over \sinh(N\kappa)}\qquad q\ge r\cr
G_{jk}(\kappa)&=G_{kj}(\kappa)=2 \coth(\kappa) \exp(-j|\kappa|)
\sinh(k\kappa)
\qquad j\ge k\cr}}

Eq. \coBAEa\ can be rewritten in a simple manner using these kernels. Let us prove for example
that $K_{2,j,k}=G_{jk}$ for $j>k$. One starts with $K_j(\kappa)=\exp(-j|\kappa|)$;
then adding the indices $k$ (resp. $2$) amounts to multiplying (for convolution) by $\sinh(k\kappa)
/\sinh(\kappa)$ (resp. $2 \cosh(\kappa)$), which produces $G_{jk}$.

Thus, Eq. \coBAEa\ becomes:
\eqn\coBAEb{\rhot^r_j+\sum_{q,k} C^{qr}\star G_{jk}\star\rho^q_k=f^r_j}
where by definition $f^r_j\equiv\delta^{r1}N_e
K_{f,j}(\Lambda-1)+\delta^{rn} K_{l,j}(\Lambda)$.
Multiplying by $C_{jk}$ we finally find:
\eqn\coBAEc{\sum_k C_{jk}\star\rhot^r_k+\sum_q C^{qr}\star\rho^q_j=\delta_{jf}\delta^{r1}N_e s(\Lambda-1)
+\delta_{jl}\delta^{rn} s(\Lambda)}

\newsec{Ground state and low-lying excitations}
In this section, we discuss the nature of the ground state and of the physical
excitations above the ground state. 
Before proceeding, the following remark should be made:

In order to get a simple consistent picture
(that sheds some light on the thermodynamic results we find in the
next sections),
we shall intentionally choose to ignore some purely discrete effects
which are connected with the exact number of electrons. As an example,
let us remember that even in the simplest Bethe Ansatz-solvable model,
the $SU(2)$ spin $1/2$ XXX model, the ground state representation
depends on the parity of the number of
spins; but the thermodynamic properties of the model do not depend
on it, and one can make the simplifying assumption that it is even.
In our case, we shall ignore these effects by allowing the $M^r$ to be
sometimes non-integer. Such a procedure should certainly not affect the
thermodynamic quantities such as the free energy in the
$L\rightarrow\infty$ limit\foot{Note that these effects {\it will}
modify the $1/L$ corrections to the free energy, which are themselves
of the same order as the free energy of the impurity (cf Eq. \Fdec\ of next
section); but the latter should be unaffected at leading order.}.

\subsec{Energy.} According to \en, 
the energy (without any magnetic field) is given by:
\eqn\ena{\eqalign{
E&=-D\sum_{\gamma=1}^{M^1} \Theta_f(\Lambda_\gamma^1-1)\cr
&=-D\sum_j\sum_{\gamma=1}^{M^1_j} \Theta_{f,j}(\Lambda^1_{j;\gamma}-1)\cr
&=-D\sum_j \int d\Lambda\, \rho^1_j(\Lambda) \Theta_{f,j}(\Lambda-1).\cr
}}
($D=N_e/L$). For future use we shall rewrite this
\eqn\enaa{E=\sum_{j,r}\int d\Lambda\, \rho^r_j(\Lambda) g^r_j(\Lambda)}
where $g^r_j(\Lambda)\equiv -D \delta^{r1} \Theta_{f,j}(\Lambda-1)$.

The energy can also be expressed in terms of $\rhot$:
\eqn\enb{\eqalign{
E&=-D\sum_{r,j,k} \int d\Lambda\, C_{jk}\star G^{r1}\star(-\rhot^r_k+f^r_k)
\Theta_{f,j}(\Lambda-1)\cr
&=\Egs+D\sum_r \int d\Lambda\, \rhot^r_f(\Lambda)
G^{r1} \star \left[2 \arctan(e^{\pi(\Lambda-1)/c})\right].\cr
}}
which we can again rewrite under the form
\eqn\enbb{E=\Egs+\sum_{r,j} \int d\Lambda\,
\rhot^r_j(\Lambda) \tilde{g}^r_j(\Lambda)}
where explicitly:
\eqn\enbc{\tilde{g}^r_j\equiv D \delta_{jf} 
\left[2\arctan\left(
\tan\left({\pi\over 2}{N-r\over N}\right) 
\tanh\left({\pi\over Nc}(\Lambda-1)\right)\right)
+\pi{N-r\over N}\right]}

We first give a brief explanation of \enb--\enbb.
$\Egs$ is the energy of the ground state. It is obtained for
$\rhot^r_f=0$,
that is the ground state is filled with the maximum amount of
$f$-strings; we shall build it explicitly in {\it 2.3}.
Above the ground state,
excitations are created by holes in the sea
of $f$-strings; the other strings do not contribute to the energy.
Their role should become transparent in next subsection.

Let us also see what happens in the scaling limit,
when the density of
electrons per unit length $D$, which will eventually
become the UV cutoff of our theory, is sent
to infinity; then, a typical state of our
system will have densities per unit length
of $f$-strings $M^r_f/L$ that will be of order $D$ and
will diverge. On the
other hand, densities of {\it holes} of $f$-strings $\tilde{M}^r_f/L$,
and of $j$-strings $M^r_j/L$
($j\ne f$) will remain of the order of the physical energy scale.
We have already rewritten the energy $E-\Egs$ in terms of the
$\rhot_f^r$ only, whose densities remain finite. It is interesting do
the same thing for the quantum numbers characterizing the $SU(N)$
representation of the state.

\subsec{Representation of a state.} We have seen that a Bethe Ansatz
state is in the representation $R=R_0-\sum_r M^r \alpha^r$, where
the $M^r$ are the numbers of roots of type $r$: $M^r=\sum_j j\, M^r_j$.
A better way to describe the corresponding $SU(N)$ Young tableau
is to count the number of columns $n^r$
of a given length $r$ ($1\le r\le N-1$). One can easily show the following
relation (cf appendix A for more details):
\eqn\qn{
n^r=\delta^{r1}N_e f+\delta^{rn} l - 2 \sum_{j,q} C^{qr} j\, M^q_j}
Here $C^{qr}$ stands for $C^{qr}(\kappa=0)$, that is (up to a factor of 2)
the usual Cartan matrix of $SU(N)$.

We shall now relate the quantum numbers $n^r$
to the numbers of holes $\tilde{M}^r_f=\int \rhot^r_f$
and of strings $M^r_j$ ($j\ne f$).

We start from equation \coBAEb\ 
with $j=f$, which we integrate from $-\infty$ to
$+\infty$ (i.e. take $\kappa=0$):
\eqn\qnBAE{\tilde{M}^r_f + \sum_q C^{qr} G_{ff} M^q_f + \sum_{k\ne f,q}
C^{qr} G_{kf} M^q_k=G_{ff} \delta^{r1} {N_e \over 2} + G_{fl}
\delta^{rn} {1\over 2}}
where the argument $\kappa=0$ is implied for all kernels. Using this
formula,
one can easily compute the sum of eq. \qn. We give the final result:
\eqn\qnf{n^r=\left\{\eqalign{
&\tilde{M}^r_f - \sum_{j>f,q} (j-f) 2\, C^{qr} M^q_j \qquad l\le f\cr
&\tilde{M}^r_f - \sum_{j>f,q} (j-f) 2\, C^{qr} M^q_j +(l-f) \delta^{rn}
\qquad l\ge f\cr
}\right.}
The first observation is that we must treat separately
the underscreened case ($l\ge f$)
and the overscreened case ($l\le f$): this is directly related to
the representation/degeneracy of the ground state, which we shall
discuss in next subsection.

Notice next that the role of the 
$j$-strings, $j>f$, which did not contribute to the energy,
is now clear: they allow to lower the spin of the system at fixed
energy, i.e. fixed physical excitations\foot{It is known that
for a small number of
excitations, the complex roots (non-$f$-strings) do not necessarily form
exact $j$-strings: the spacing of their imaginary parts might deviate from
the string behavior. However this does not modify qualitative analysis.}.
This is the usual way, in
Bethe Ansatz systems, to select between the different irreducible
subrepresentations inside the tensor product of the representations
of the physical excitations (without any $j$-strings, $j>f$, the
system is in the highest possible weight representation).

On the other hand the $j$-strings ($j<f$) do not change the
representation of the state: in fact they play a role which is
quite similar to the $j$-strings ($j>f$), but for
an other quantum number which can be associated with a state,
its $SU_q(N)$ representation
($q=\exp(2i\pi/(f+N))$, so that
the set of representations is ``restricted to level $f$''). 
We shall elaborate on this new quantum number in section 5.

\subsec{The ground state.}
According to Eq. \enb, the ground state is obtained for $\rhot^r_f=0$.
This means that the ground state is filled with a continuous density
of $f$-strings of all types $r=1\ldots N-1$. It can then be shown
that there are no $j$-strings ($j\ne f$); this allows to calculate the
densities $\rho^r_f$ from \coBAEb:
\eqn\gsr{\rho^r_f(\Lambda)=N_e G^{r1}\star s(\Lambda-1)+G^{rn}
\star(G_{ff})^{-1}\star K_{lf}(\Lambda)}
We shall now discuss separately the different cases:

\blob $f\le l$ (underscreening). We first compute the ground state
representation.
Applying \qnf\ to the ground
state, we find the following Young tableau (fig.~\und).
\fig\und{Underscreening of the impurity by the electrons. On this
example, $3$ fused $f=2$ electrons screen the $5\times 4$ impurity,
reducing its spin to $5\times 2$.}{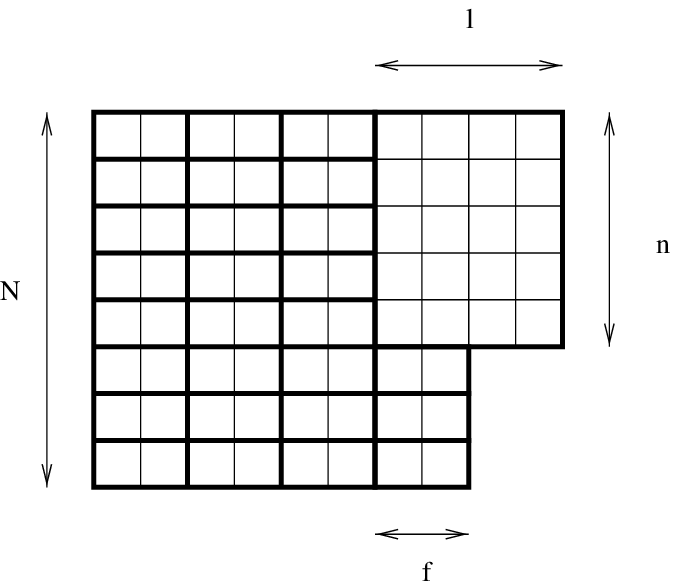} 
Let us try to interpret intuitively this Young tableau. First there
are a large number (of order $N_e$) of electrons that are not directly
interacting with the impurity: they are represented by the ``trivial''
part (i.e. $N$ rows) of the Young tableau, to the left of the impurity.
Then there are exactly $N-n$ fused electrons which can be thought as
on the same site as or glued to the impurity:
Their effect is to reduce
its spin from $n\times l$ to $n\times (l-f)$. To confirm this
analysis, one can calculate the density of holes of $j$-strings
$\rhot^r_j$; one finds:
\eqn\strhol{
\rhot^r_j(\Lambda)=\delta^{rn}(K_{l,j}(\Lambda)
-G_{jf}\star (G_{ff})^{-1} K_{f,l}(\Lambda))
}
For $j>f$, $\rhot^n_j\ne 0$ so there are holes of $j$-strings; 
but one can check that there
are ``not enough'' to actually create $j$-strings (this confirms the
fact that the ground state cannot contain $j$-strings). The
interpretation of this result is the following: the underscreened
impurity is in a given irreducible representation of $SU(N)$, so its
spin cannot be changed. However, if one adds one physical excitation into
the system, it will create more holes of $j$-strings and allow creation
of $j$-strings: that is so because the tensor product of the
(non trivial) representation of the impurity and of another
non trivial representation is
necessarily reducible, and the choice between the different
irreducible subrepresentations is made, as has already been mentioned,
by inserting appropriate $j$-strings (cf Eq. \qnf).

On the other hand, for $j<f$, one finds from \strhol\ that
$\rhot^r_j=0$: the screened impurity possesses a trivial $SU_q(N)$ quantum
number.

\blob $f\ge l$ (overscreening). This time Eq. \qnf\ implies that
the ground state is in the trivial representation, which is not what
one might naively expect; indeed, by minimizing only the interaction
part $J
\bar{\psi}^a_i(0) \sigma_A^{ab} \psi^b_i(0) 
\bar{\chi}^{\alpha}\tau_A^{
\alpha \beta}\chi^{\beta}$ of the Hamiltonian, one would obtain a different ground state
representation~\ref\AJ{A.~Jerez,N.~Andrei  and
G.~Zarand, in preparation.}. Let us mention now that
what this proves is that
the RG fixed point $J=\infty$ is unstable to the kinetic part of the
Hamiltonian and that we have a non-trivial IR fixed point $J=J^\star$.
This will be explained more carefully
when we study the thermodynamics of the model.

One can again calculate the densities of holes of $j$-strings:
the formula \strhol\ is still valid; but the analysis is reversed.
For $j>f$, one find $\rhot^r_j=0$, which confirms the fact that
the ground state is in the trivial representation.
On the contrary we now have holes of $j$-strings for $j<f$:
the impurity possesses a non-trivial $SU_q(N)$ quantum number,
though it originally possessed none. Of course its origin is related
to the presence of the electrons that screen the impurity.
We shall discuss this in detail when we consider
the entropy at zero temperature.

\subsec{Physical excitations.}
We have already mentioned that physical spin excitations are created
by inserting holes of $f$-strings. As there are $N-1$ types of holes,
we conclude that there are $N-1$ types of physical particles labeled
by $r=1\ldots N-1$.
The rapidity $\Lambda$ at which the
hole is inserted determines the energy and momentum of the corresponding
excitation. We have already given in {\it 2.1} the energy
$\epsilon^r(\Lambda)=\tilde{g}^r_f(\Lambda)$: (Eq. \enbc)
\eqn\enbd{\epsilon^r(\Lambda)=D
\left[2\arctan\left(
\tan\left({\pi\over 2}{N-r\over N}\right) 
\tanh\left({\pi\over Nc}(\Lambda-1)\right)\right)
+\pi{N-r\over N}\right]}
As $\epsilon^r(-\infty)=0$, these are massless excitations. From
general arguments it is clear that they have a linear dispersion
relation, so that their momentum $p^r=\epsilon^r$. This can also be easily
extracted from the Bethe Ansatz equations by writing a phase shift
condition on a compactified space for one single physical excitation.

We may at this point
go to the scaling limit $D\rightarrow\infty$, $c\rightarrow 0$
keeping the physical scale $T_0\equiv D e^{-2\pi/Nc}$ fixed.
The energy/momentum then takes the simple form: 
\eqn\scalep{\epsilon^r=p^r=2T_0 \sin\left(\pi r\over N\right) 
e^{2\pi\Lambda/Nc}}
which is characteristic of a relativistic (massless) right-moving
particle, with $2\pi\Lambda/Nc$ its rapidity.

We also know their representation from {\it 2.2}: according to Eq.
\qnf, inserting a hole of type $r$ creates a column of size $r$ in
the corresponding Young tableau; that is, the particle of
type $r$ belongs to the fundamental (or totally antisymmetric)
representation of $SU(N)$ with $r$ boxes. It can also be shown
that they belong to the same representation of the quantum group
$SU_q(N)$.

These particles interact with each other and with the impurity, which
leads to the concept of phaseshift. The latter can be extracted from
the Bethe Ansatz equations; however, we
postpone their detailed study to a future publication.

\newsec{Thermodynamic Bethe Ansatz equations (T.B.A.)}
We now want to study the model at finite temperature $T$ ($T\gg 1/L$).
In the standard way, one derives the T.B.A. by minimizing
the free energy $F=E-TS$ with respect to $\rho^r_j$; using \coBAEb\ one finds
that $\delta S/\delta\rho^r_j$ equals:
($\eta^r_j\equiv\rhot^r_j/\rho^r_j$)
\eqn\TBA{\log(1+\eta^r_j)-\sum_{q,k} C^{qr}\star G_{jk}
\star\log(1+(\eta^q_k)^{-1})={g^r_j\over T}}
where we have used $g^r_j=\delta E/\delta\rho^r_j$, 
the energy of a $j$-string of type $r$ (Eq. \enaa).
Note that the T.B.A. do not depend on the representation of the impurity.
In fact one can say that the T.B.A. only describe the electrons (in a way
appropriate for studying their interaction with the impurity) and not
the impurity itself. On the contrary, the fusion equations that will
be written later describe specifically the impurity.

\subsec{Free energy.} General T.B.A. formulae imply that
\eqn\FTBA{\eqalign{
F&=-T\sum_{j,r} \int d\Lambda\, f^r_j(\Lambda) 
\log(1+(\eta^r_j(\Lambda))^{-1})\cr
\hbox{\rm (using T.B.A.)}\, &=-\sum_{j,k,q,r} \int d\Lambda\, G^{qr}\star
C_{jk} \star f^q_k(\Lambda)
\left[-g^r_j(\Lambda)+T\log(1+\eta^r_j(\Lambda))\right]\cr
}}
The first term is the ground state energy.
In the second term, the explicit expression of $f^q_k$ leads to
\eqn\Fdec{
F=\Egs+L{\cal F}+F^n_l,}
where ${\cal F}$ is the free energy per unit length of the electrons:
\eqn\Fe{{\cal F}=-D T \sum_r \int d\Lambda\, \Ghat^{r1}
(\Lambda-1) \log(1+\eta^r_f(\Lambda))}
and $F^n_l$ is the free energy of the impurity:
\eqn\Fi{F^n_l=-T\sum_r \int d\Lambda\, \Ghat^{rn}(\Lambda) 
\log(1+\eta^r_l(\Lambda)).}
Here we have introduced a new notation which will prove convenient:
$\Ghat^{qr}(\Lambda)\equiv G^{qr}\star s(\Lambda)$; $\Ghat^{qr}$ has
the following Fourier transform:
\eqn\Ghatf{\Ghat^{qr}(\kappa)=\Ghat^{rq}(\kappa)=
{\sinh((N-q)\kappa)\sinh(r\kappa)\over\sinh(\kappa)\sinh(N\kappa)}
\qquad q\ge r}

\subsec{Scaling limit.} We now take the limit $c\rightarrow 0$,
$D\rightarrow\infty$, keeping $T_0\equiv D e^{-2\pi/Nc}$ fixed; $T_0$
is the dynamically generated energy scale due to the presence of
the impurity. We rewrite the T.B.A. equations \TBA\ under the form
\eqn\TBAb{\log(1+(\eta^r_j)^{-1})
-\sum_{q,k} G^{qr}\star C_{jk}\star \log(1+\eta^q_k) 
= {\tilde{g}^r_j\over T}}
which become in the scaling limit (Eq. \scalep)
\eqn\scTBA{\log(1+(\eta^r_j)^{-1})
-\sum_{q,k} G^{qr}\star C_{jk}\star \log(1+\eta^q_k) 
= 2\delta_{jf}
\sin\left({\pi r\over N}\right)
e^\zeta}
where we have successively rescaled\foot{Note
that the rescaling also induces a rescaling of the kernels,
according to the rule $\phi(\zeta)d\zeta=\phi(\Lambda)d\Lambda$.}
and shifted the rapidity:
\eqn\resc{\zeta={2\pi\Lambda\over Nc}+\log(T_0/T).}

The shift removes any
dependence of the T.B.A. on $T$. The qualitative picture of the
behavior of these equations
is then the following: there are two asymptotic regimes
$\zeta\rightarrow\pm\infty$ characterized by limiting values
$\eta^r_j(\pm\infty)$. We shall derive these values in the next subsection,
once we have explained their group-theoretic meaning.

As the crossover between the two regimes occurs for $\zeta\approx 0$
that is ${2\pi\over Nc}\Lambda\approx\log(T/T_0)$,
$\zeta\rightarrow -\infty$ will be called the
high-temperature regime, whereas $\zeta\rightarrow +\infty$ will be the
low-temperature regime.

Let us also rewrite the free energy in terms of the new variable:
we use for example $\Ghat^{r1}(\Lambda)=
(1/Nc) \sin(\pi r/N)/(\cosh((2\pi/Nc)\Lambda)-\cos(\pi r/N))$:
\eqn\Fscal{\eqalign{
{\cal F}&=-{T^2\over\pi} \sum_r \sin\left({\pi r\over N}\right)
\int d\zeta \, e^\zeta \, \log(1+\eta^r_f(\zeta))\cr
F^n_l&=-T\sum_r \int d\zeta \, \Ghat^{rn}(\zeta-\log(T_0/T))
\log(1+\eta^r_l(\zeta))\cr
}}
The fact that $T_0$ only appears in the free energy of the impurity
indicates that it is the presence of the impurity which triggered
the energy scale dynamical generation.

\subsec{Fusion equations and interpretation of limiting values.}
We shall now work with the T.B.A. equations \scTBA, i.e. after the scaling
limit has been taken. Let us see how these equations are related
to fusion equations for rectangular Young tableaux.
The first step is to analytically continue
the functions $\eta^r_j$ to the strip $|\Im\zeta|\le \pi/N$, so that
they possess no zeros or poles for $|\Im\zeta|<\pi/N$. Then
$\log(1+\eta^r_j)$ and $\log(1+(\eta^r_j)^{-1}$ can also be continued
in the same way. For any function that satisfies this analyticity
requirement one can then define an inverse operator of $\Ghat^{qr}$
which we of course denote by $\Chat^{qr}$; this operator is {\it not} a
convolution kernel since it acts in the following way:
\eqn\defChat{\Chat^{qr}\star\phi(\zeta)=\delta^{qr}
(\phi(\zeta+i\pi/N)+\phi(\zeta-i\pi/N))-(\delta^{qr+1}+\delta^{qr-1})
\phi(\zeta)}
$\Chat^{qr}$ can be considered as a discrete Laplace operator in
$(r,\zeta)$ space. As $\Ghat^{qr}=G^{qr}\star s$, one should have
$C^{qr}=\Chat^{qr}\star s$: this is indeed the case on condition that
one correctly remains inside the analyticity strip when performing
the convolution integral (fig. \strip).
\fig\strip{$s(\zeta)=N/(4\pi\cosh(N\zeta/2))$
has poles at $\zeta=\pm i\pi/N$, so that
when one considers the convolution kernel $s(\zeta+i\pi/N)+
s(\zeta-i\pi/N)$, one should deform the integration contour so as to
remain inside the analyticity strip. Then due to
$2i\pi/N$-antiperiodicity only the pole contribution at
$\zeta=+i\pi/N$ remains, so that
$s(\zeta+i\pi/N)+s(\zeta-i\pi/N)=\delta(\zeta)$.}{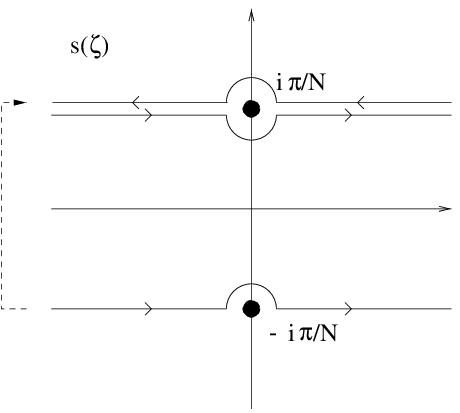} 

We now take eq. \scTBA\ and act on it 
with $\Chat^{qr}$: this annihilates the r.h.s. and we find after
exponentiation:
\eqn\fusa{\eta^r_j(\zeta+i\pi/N) \eta^r_j(\zeta-i\pi/N)=
{\left[1+\eta^r_{j+1}(\zeta)\right] \left[1+\eta^r_{j-1}(\zeta)\right]
\over \left[1+(\eta^{r+1}_j(\zeta))^{-1}\right]
\left[1+(\eta^{r-1}_j(\zeta))^{-1}\right]}.}
This system of equations 
(``$Y\!$-system'')~\ref\Z{Al.~B.~Zamolodchikov, {\it Phys. Lett.} B253
(1991), 391.}
is well-known 
to be related to $SU(N)$ ``spectral parameter-dependent'' 
fusion equations~\ref\Yfus{The connection between $Y\!$-system
and fusion equations for the $SU(2)$ case is implicit in
A.~Kl\"umper and P.~Pearce, {\it Physica} A183 (1992), 304\semi
and stated in
A.~Kuniba
and T.~Nakanishi, {\it Mod. Phys. Lett.} A7 (1992), 3487\semi
see also 
V.~V.~Bazhanov and N.~Reshetikhin, {\it J. Phys} A23 (1990), 1477.
}.
For now, we simply note the ``rank-level duality'' of these equations:
$j\leftrightarrow r$, $\eta\leftrightarrow \eta^{-1}$.
In fact, we have to be careful because of the non-trivial r.h.s. of
Eq. \scTBA, which does not appear in the framework of normal fusion
equations. 
We shall therefore
proceed with caution and rederive the fusion equations from
the very beginning. First we introduce the generalized characters
\eqn\defch{\chi^r_j(\zeta)\equiv 
\exp \left[ \sum_q \Ghat^{qr} \star \log (1+\eta^q_j) \right]
}
(we call them generalized characters
because normal characters
are the solutions of the standard
fusion equations, whereas here we deal with modified fusion equations). 
Notice the similarity with Eq. \Fi\ or \Fscal: one has
\eqn\chF{\log\chi^n_l(\zeta)=-{1\over T}F^n_l(T=T_0 e^{-\zeta})}
In fact, as the T.B.A. equations
do not depend on the representation of the impurity, one can consider
that the $\chi^r_j$, for {\it any} values of $r$ and $j$, are related
to the free energy $F^r_j$ of an impurity in the representation $r\times j$
($r$ rows, $j$ columns)~!

One can invert relation \defch\ using the definition of $\Chat^{qr}$:
\eqn\defchb{1+\eta^r_j(\zeta) =
{\chi^r_j(\zeta+i\pi/N) \chi^r_j(\zeta-i\pi/N)
\over \chi^{r+1}_j(\zeta) \chi^{r-1}_j(\zeta)}}
($\chi^0_j$ and $\chi^N_j$ are by convention equal to $1$).
So far we have not used the T.B.A. \scTBA\ yet. We now do so
in order to derive ``dual'' (in the sense of
rank-level duality) expressions for $1+(\eta^r_j)^{-1}$.
First we introduce a $\Chat_{jk}$ defined by the same formula
\defChat, and such that $\Chat_{jk}\star s=C_{jk}$.
As $G^{qr}\star C_{jk}=\Ghat^{qr}\star \Chat_{jk}$, the
insertion of the definition of the characters in \scTBA\ gives:
\eqn\dualch{\log(1+(\eta^r_j)^{-1}) = \sum_k \Chat_{jk}\star \log \chi^r_k
+ 2\delta_{jf} \sin\left({\pi r\over N}\right) e^\zeta}
Notice the additional term due to the r.h.s. of \scTBA, 
which leads to the following modified formula:
\eqn\dualchb{1+(\eta^r_j(\zeta))^{-1}=
{\chi^r_j(\zeta+i\pi/N) \chi^r_j(\zeta-i\pi/N)
\over \chi^r_{j+1}(\zeta) \chi^r_{j-1}(\zeta)} 
e^{2\delta_{jf} \sin(\pi r/N) e^\zeta}}
($\chi^r_0\equiv 1$) or finally
\eqn\dualchc{\eta^r_j(\zeta)={\chi^r_{j+1}(\zeta) \chi^r_{j-1}(\zeta)
\over \chi^{r+1}_j(\zeta) \chi^{r-1}_j(\zeta)}
e^{-2\delta_{jf} \sin(\pi r/N) e^\zeta}.}
This, with the boundary conditions $\chi^0_j=\chi^N_j=1$
and $\chi^r_0=1$, allows to compute $\eta^r_j(\zeta)$ as a function of
the $\chi^t_k(\zeta)$.
Remembering that $\chi^t_k$ is directly related to the free energy
of an impurity in the representation $t\times k$, one
can consider that one puts ``test'' impurities (corresponding to
any rectangular Young tableau) in the system in order
to find the characters $\chi^r_j$, and from there the functions $\eta^r_j$.

One can now proceed to write down modified fusion equations
for the $\chi^r_j$: rewriting the trivial identity
$\defchb=1+\dualchc$ we find
\eqn\fus{\chi^r_j(\zeta+i\pi/N) \chi^r_j(\zeta-i\pi/N)
=\chi^{r+1}_j(\zeta) \chi^{r-1}_j(\zeta) 
+ \chi^r_{j+1}(\zeta) \chi^r_{j-1}(\zeta) 
e^{-2\delta_{jf} \sin(\pi r/N) e^\zeta}.}
This system of coupled non-linear equations can be represented by 
fig.~\fuspic.
\fig\fuspic{Pictorial representation of the modified fusion equations.
The marked circles correspond to the position of the
modified term. The square around the circle $n,l$ means that it is
$\chi^n_l$ which gives
the free energy of the impurity.}{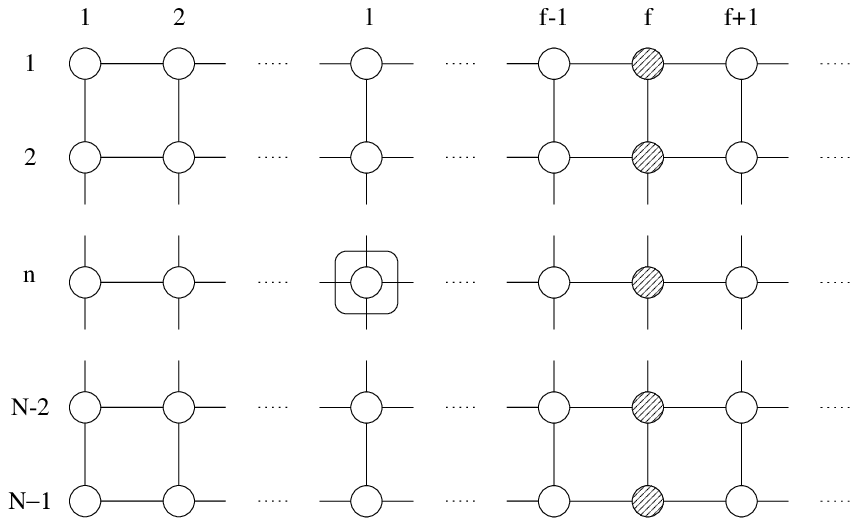} 
Just like the T.B.A. equations, the fusion equations contain all
the physics of the Kondo model. In fact, it is the additional term
$\exp(-2\delta_{jf} \sin(\pi r/N) e^\zeta)$ which really contains
all the information on the crossover from the low-temperature
regime to the high-temperature regime. We shall elaborate on this
in next section which contains all the calculations of thermodynamic
quantities.

For the moment, in order to see the role of the additional term
in a very simple setting, let us consider
the limits $\zeta\rightarrow\pm\infty$; then the shifts of
$\pm i\pi/N$ of $\zeta$ become negligible, which means that
the equations \fus\
reduce to $SU(N)$ tensor product equations for rectangular
Young tableaux, and the $\chi^r_j$ tend to
asymptotic values $\chi^r_j(\pm\infty)$
which should be ordinary $SU(N)$ characters. We
now compute these values in the absence of magnetic field:

\blob
When $\zeta\rightarrow -\infty$, the modification induced by the r.h.s.
of \scTBA\ vanishes and one simply obtains ordinary tensor product equations
for any $j\ge 1$ and $r$. Therefore
\eqn\fushiT{\chi^r_j(-\infty)=\chi^r_j(G)}
where $\chi^r_j(G)$ is the character associated with some matrix $G\in SU(N)$
and the rectangular Young tableau $r\times j$ (cf appendix A for a definition).

The constraint that $\eta^r_j>0$ (i.e. $\chi^r_j>0$)
for all $r$ and $j$, the boundary
conditions $\chi^r_0=1$, $\chi^r_{-1}=0$ and
the asymptotic behavior of $\chi^r_j$ as $j\rightarrow\infty$ (we shall
work out the latter in the more general case of the presence
of a magnetic field, see {\it 3.4})
lead to the unique
solution $G=1_{SU(N)}$, i.e. the T.B.A. select the Perron-Frobenius eigenvalue
of the tensor product matrix corresponding to the Young tableau $r\times j$.
This gives the following values of $\eta^r_j(-\infty)$:
\eqn\asyhiT{\eta^r_j(-\infty)={j(j+N)\over r(N-r).}}

\blob
When $\zeta\rightarrow +\infty$, the fusion
equations \fus\ are cut in two separate parts: for $j\le f$, 
we can consider that we have normal fusion equations with
the additional boundary condition
$\chi^r_{f+1}=0$; this, with the usual boundary conditions $\chi^r_0=1$,
$\chi^r_{-1}=0$, implies that
\eqn\fusloTa{\chi^r_j(+\infty)=\chi^r_j(G_f) \qquad j\le f}
where $G_f$ is a $SU(N)$ matrix which can be uniquely determined
by imposing $\eta^r_j>0$: ($\theta\equiv \pi/(f+N)$)
\eqn\fusloTb{G_f=\left( \matrix{
e^{i(-N+1) \theta} &&&&0\cr
&e^{i(-N+3)\theta} &&&\cr
&&\ddots &&\cr
&&& e^{i(N-3) \theta} &\cr
0&&&& e^{i(N-1) \theta}\cr
}\right)}
We shall see in section 5 that this character can again
be interpreted as the Perron-Frobenius
eigenvalue of an appropriately truncated (at level $f$) tensor product matrix.

For $j>f$, this time everything happens
as if $\chi^r_{f-1}=0$. Together with $\chi^r_f=1$ (this comes out of
\fusloTa-\fusloTb) and the asymptotic
behavior of $\chi^r_j$ as $j\rightarrow\infty$, it reproduces the
boundary conditions found in the limit $\zeta\rightarrow -\infty$, except
for a shift of $f$ in the string index. Therefore:
\eqn\fusloTc{\chi^r_j(+\infty)=\chi^r_{j-f}(1_{SU(N)}) \qquad j>f}
Finally,
\eqn\asyloT{\eta^r_j(+\infty)=\left\{\eqalign{
&{\sin(j\theta) \sin((j+N)\theta) \over \sin(r\theta) \sin((N-r)\theta)}
\qquad 1\le j \le f\cr
&{(j-f)(j-f+N)\over r(N-r)}\qquad j\ge f\cr
}\right.}

\subsec{Magnetic field.} 
In a $SU(N)$-invariant model,
the most general magnetic field $B$ one can impose
is an arbitrary element of the Lie algebra $\goth{su}(N)$:
\eqn\genmag{B=\sum_{A=1}^{N^2-1} B_A T_A}
where the $T_A$ are the generators of $\goth{su}(N)$ acting on the whole
Hilbert space of our model. Using the $SU(N)$ symmetry one can suppose
that $B$ belongs to the Cartan subalgebra;
this means that $B$
can be described in the fundamental representation of $SU(N)$
as the diagonal matrix:
\eqn\diagmag{B=\left( \matrix{B^1 &     &      &  0  \cr
                       & B^2 &      &     \cr
                       &     &\ddots&     \cr
                    0  &     &      & B^N \cr
}\right)}
with the condition $B^1+B^2+\cdots+B^N=0$.
One still has a residual symmetry of permutation of the eigenvalues of $B$
which allows to choose: $B^1<B^2<\ldots <B^N$. This choice will prove
convenient later. Of course in the $SU(2)$ case $B$ has only two
eigenvalues $B^1\equiv -B$ and $B^2=B$, and we recover the usual
one-component magnetic field.

The effect of the magnetic field is that, for a given Bethe Ansatz
solution belonging to an irreducible
representation $R$ (characterized by its highest
weights, or by the numbers $M^r$),
the states with different $SU(N)$ quantum numbers (weights)
have a different energy. In other words, the corresponding
energy level $E$ has been split in several levels; 
the resulting contribution to the partition function then
factorizes: $Z=Z_B Z_{\cal H}$, with $Z_{\cal H}=\exp(-E/T)$ and
\eqn\Zmag{Z_B=\chi_R\bigl(e^{-B/T}\bigr)}
where $e^{-B/T}$ is the diagonal matrix with eigenvalues $e^{-B^a/T}$,
and $\chi_R$ denotes the character of the representation $R$.
Fortunately, this expression simplifies in the thermodynamic limit,
when one deals with large Young tableaux: then, one can show (see
appendix A) that
\eqn\chiasy{Z_B \propto e^{-\sum_{r=1}^{N-1} M^r b^r/T}}
Here one has used the ordering of the $B^a$, and defined $b^r=B^{r+1}-B^r$.
Hence, $-b^r$ effectively acts as a {\it chemical potential} for the
$r^{\rm th}$ type of B.A. roots\foot{Actually, 
even for $B=0$, $Z_B$ gives an additional entropic
factor in the partition function
that we have not taken into account so far. But its effect
is negligible in the thermodynamic limit: only the leading
behavior embodied in Eq. \chiasy\ matters.}.

The change induced by the addition of a magnetic field in the T.B.A. is
straightforward: the r.h.s. of \TBA\ is now:
\eqn\TBAmag{{g^r_j\over T}=-{D\over T} \delta^{r1}
\Theta_{f,j}(\Lambda-1)+j{b^r\over T}}

The other equations \FTBA--\fus\ are unchanged: the presence of the
magnetic field is hidden in the asymptotics of $\eta^r_j$ as
$j\rightarrow\infty$. Starting from \TBA, simple manipulations show
that
\eqn\asyj{\lim_{j\rightarrow\infty} 
K_j\star\log(1+\eta^r_{j+1})-K_{j+1}\star
\log(1+\eta^r_j)={b^r\over T}}
or for the characters
\eqn\asyjb{\lim_{j\rightarrow\infty} 
K_j\star\log\chi^r_{j+1}-K_{j+1}\star\log\chi^r_j
=\sum_{a=1}^r {B^a\over T}.}

\newsec{Calculation of thermodynamic quantities.}
We shall work directly with the $\chi^r_j$ and the fusion equations 
\fus\ they satisfy, since they will provide us with the free
energy of the impurity, through the
correspondence \chF\ that we recall:
\eqn\chFb{F^n_l(T)=-T \log\chi^n_l(\zeta=\log(T_0/T)).}

\subsec{High temperature ($T\gg T_0$).}
The high temperature behavior is governed by
the $\zeta\rightarrow -\infty$ region of the T.B.A.
In this region, the modified fusion equations become ``free'' in the
sense that the extra term of Eqs. \fus\ $\exp(-2\delta_{jf}
\sin(\pi r/N) T_0/T)=1$ up to $1/T$ corrections.
Let us see what this means physically for the impurity:
 
When a magnetic field is present, one can show that the
solution \fushiT\ of the fusion equations is still valid,
except that the matrix
$G$ is no more restricted to be in $SU(N)$ (only in $SL(N)$); in
fact, the choice
\eqn\fushiTB{G=e^{-B/T}}
gives the correct asymptotics \asyjb, as can be checked by
applying the asymptotic form of characters 
for large Young tableaux (appendix A)
to $\chi^r_j$, $j\rightarrow\infty$. There are $1/\zeta$
corrections\foot{These power-law corrections are characteristic of a system
of T.B.A. equations for which the index $j$ takes an {\it infinite}
set of values.}
to the $\zeta\rightarrow -\infty$ behavior
of the $\chi^r_j$,
which gives the following expansion:
\eqn\asyFhiT{F^n_l\buildrel T\rightarrow\infty\over = -T
\left[ \log\chi^n_l\bigl(e^{-B/T}\bigr)
+ {\alpha^n_l\over\log(T/T_0)}+\ldots\right]} 
The first term is simply, as already noted (Eq. \Zmag) the
free energy of a spin in the representation $n\times l$, that
is the impurity without any interaction with the electrons.
This
indicates a $J=0$ UV fixed point, with logarithmic corrections
characteristic of asymptotic freedom.

\subsec{Low temperature ($T\ll T_0$, $B\lesssim T$).}
The region where both $T$ and $B$ are small compared to $T_0$
corresponds to the vicinity of $\zeta=+\infty$.

We shall in particular
consider $F^n_l$ for very low magnetic fields ($B\ll T)$;
due to $SU(N)$-invariance the expansion takes the form (cf appendix A):
\eqn\defsusc{F^n_l
\buildrel B\rightarrow 0\over \sim
\sigma^n_l {1\over 2N}\sum_{a=1}^N (B^a)^2
}
which defines the magnetic susceptibility $\sigma^n_l$.

The extra term of Eq. \fus\ is now exponentially small, so that
one obtains a separation of the equations
according to $j<f$, $j=f$, $j>f$. Note that to calculate the
corrections to the $\zeta=+\infty$ value of $\chi^n_l$, we do need
to take into account the shifts of $\pm i\pi/N$ of \fus.

\blob Underscreened case ($f<l$).
Again, it
is easy to extend the $\zeta\rightarrow +\infty$ limit
\fusloTc\ to include a magnetic field; taking into account the power-law
corrections, one finds that
\eqn\asyFloTu{F^n_l\buildrel T\rightarrow 0\over = -T
\left[ \log\chi^n_{l-f}\bigl(e^{-B/T}\bigr)
 + {\beta^n_l\over\log(T/T_0)}+\ldots\right]}
The first term is interpreted by saying that the impurity
has been (under)screened by the electrons (cf fig.~\und), and
it now behaves like a spin in the $n\times(l-f)$ representation.
This indicates a $J=\infty$ IR fixed point, the logarithmic
corrections correspond to IR asymptotic freedom.

In particular, since the electrons do not contribute to the
zero temperature and zero magnetic field entropy
(cf appendix B), we can extract the latter from Eq. \asyFloTu\ and 
check that it is just the logarithm of the dimension of
the representation of the ground state, as expected.

For the susceptibility one finds
\eqn\asyFloTBu{\sigma^n_l
\buildrel T\rightarrow 0\over \sim
-{1\over T} {n(N-n)(l-f)(l-f+N)\over N^2-1}}
which contains the normal low-temperature $1/T$ divergence.

\blob Screened case ($f=l$).
The limit $\chi^n_f
(\zeta\rightarrow +\infty)= 1$ indicates that
there is no term of order $T$ in the expansion of $F^n_f$. This
corresponds to the intuitive fact that the impurity has been
completely screened by the electrons ($J=\infty$ fixed point).
Expanding around $\zeta=+\infty$ in the fusion equations for $j=f$
one finds particularly simple (linear) equations and
the expansion $\chi^n_f=1+cst\, \sin(\pi n/N)
e^{-\zeta}$ plus next corrections of the type $e^{-k\zeta}$, $k$
integer;
but not the constant in front of the first correction.
Fortunately, for $f=l$, one can use an alternative method.
One expands \Fscal\ in the limit $T\ll T_0$:
\eqn\asyFloTs{\eqalign{
F^n_f&\buildrel T\rightarrow 0\over\sim
-{T^2\over\pi T_0} \sum_r 
{\sin\left({\pi r\over N}\right)
\sin\left({\pi n\over N}\right)
\over\sin\left({\pi\over N}\right)}
\int d\zeta \, e^\zeta \, \log(1+\eta^r_f)\cr
&={1\over T_0}
{\sin\left({\pi n\over N}\right)
\over\sin\left({\pi \over N}\right)}
{\cal F}}}
so that we are led to the computation of $\cal F$, which
is done in appendix B. We only reproduce here the result \Bfinal:
\eqn\Fdilog{F^n_f
\buildrel T\rightarrow 0\over\sim
-f{\sin\left({\pi n\over N}\right)
\over\sin\left({\pi \over N}\right)}
\left[
{\pi\over 12}
{N^2-1\over N+f}
{T^2\over T_0} 
+{1\over 4\pi} \sum_{a=1}^N {(B^a)^2\over T_0}
\right]
}
Furthermore, according to the form of the next corrections
($e^{-k\zeta}$), $F^n_f$ has only integer powers of $T$ in its
expansion.

The magnetic susceptibility $\sigma^n_f$ is
\eqn\asyFloTBs{\sigma^n_f
\buildrel T\rightarrow 0\over\sim
-{1\over T_0}
{\sin\left({\pi n\over N}\right)
\over\sin\left({\pi \over N}\right)}
{Nf\over 2\pi}}
which reaches a finite limit as $T\rightarrow 0$. This confirms that the impurity
is completely screened in the IR.

\blob Overscreened case ($f>l$). 
This is the
most interesting case. Independently of the magnetic field,
one find the result \fusloTa, or more explicitly
\eqn\asyFloTo{F^n_l\buildrel T\rightarrow 0\over\sim -T\log
\left[ \prod_{p=1}^n \prod_{q=n+1}^N
{\sin(q-p+l)\theta\over\sin(q-p)\theta}\right]}
where $\theta=\pi/(f+N)$.

It is clear that here, the entropy at zero
temperature $S(T=0)=\log\chi^n_l(G_f)$
is not the logarithm of an integer number, a phenomenon
which has a simple explanation (see next section).
In particular, it is smaller than the naive value
$\chi^n_{f-l}(1_{SU(N)})$ one would obtain if one minimized only
the interaction term in the Hamiltonian \AJ. This corresponds to 
the fact that the RG flow reaches a non-trivial IR fixed point $J=J^\star$
characterized by Non-Fermi Liquid behavior.

One should note that the number $\chi^n_l(G_f)$ involved is directly
determined by the geometry of the fusion equations diagram (fig.~\fuspicb).
\fig\fuspicb{Pictorial representation of the fusion equations $j<f$ in the
limit $\zeta\rightarrow\infty$, when they
decouple from the equations $j>f$.}{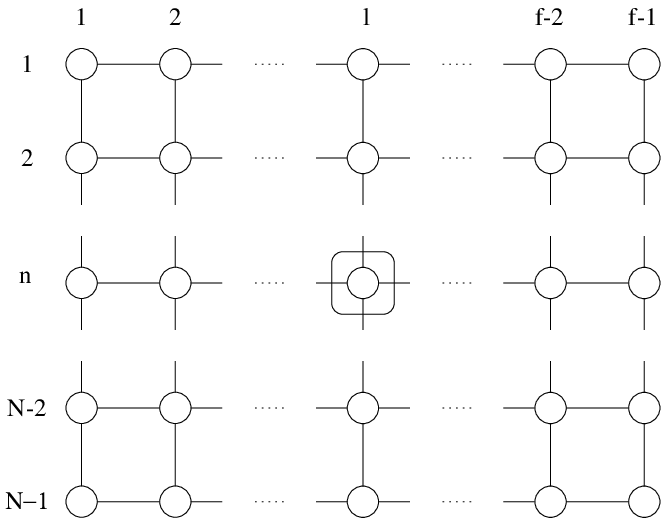} 
The symmetries of the diagram imply equalities of zero temperature
entropies for different impurities:
obviously $\chi^n_l(G_f)=\chi^{N-n}_l(G_f)=\chi^n_{f-l}(G_f)$.
Furthermore, one can see on this graphical representation
that the rank-level duality of the equations implies that
the simultaneous interchange of $n\leftrightarrow l$
and $N\leftrightarrow f$ should leave $\chi^n_l(G_f)$ unchanged.
For example, this
relates the cases of an overscreened impurity in a completely
symmetric representation ($n=1$) and in a completely antisymmetric
representation ($l=1$) (this particular duality can be found
in the large $N$ limit in~\ref\GP{O.~Parcollet
and A.~Georges, preprint cond-mat/9707337\semi
O.~Parcollet, A.~Georges, G.~Kotliar and A.~Sengupta, preprint
cond-mat/9711192.})

Let us also calculate the next correction to the free energy. Since
we are now dealing with a finite set of coupled equations ($1\le j\le f-1$,
$1\le r\le N-1$), one expects exponentially small corrections to the
dominant behavior. We already know from \Fdilog\ that 
$\chi^r_f=1+O(e^{-\zeta})$; and $\chi^r_f$ is
connected to $\chi^r_{f-1}$ by the fusion equations
(even in the $\zeta\rightarrow +\infty$ limit: take \fus\ for $j=f-1$),
and from there on to all the $\chi^r_j$, $j<f$.
This implies that all the 
$\chi^r_j$ have corrections
of order $e^{-\zeta}$; but these may be subdominant compared to other
exponentially small corrections. Hence, we try the ansatz
\eqn\anstau{\chi^r_j(\zeta)=\chi^r_j(+\infty) 
\left( 1+ a^r_j \, e^{-\tau\zeta}\right)}
for $1\le j\le f-1$, with $0<\tau<1$. Expanding the fusion equations
\fus\ around $\zeta\rightarrow +\infty$
leads to the following set of linear equations for the $a^r_j$:
\eqn\lintau{
{1\over 1+\eta^r_j(+\infty)} (a^{r+1}_j+a^{r-1}_j)
+{1\over 1+ (\eta^r_j(+\infty))^{-1}} (a^r_{j+1}+a^r_{j-1})
=\lambda a^r_j}
where $\lambda\equiv 2 \cos(\pi\tau/N)$, and we
have the boundary conditions
$a^0_j=a^N_j=a^r_0=a^r_f=0$.
This is an eigenvalue problem;
the dominant correction in \anstau\ corresponds to the biggest $\lambda$,
so we are looking for the Perron-Frobenius 
eigenvalue and eigenvector\foot{Note
that the solution $a^r_j=1$ trivially satisfies \lintau\ 
with $\lambda=2$, but not the boundary conditions.}.
The solution
\eqn\soltau{\eqalign{
a^r_j&=a\, \sin(r\theta)\sin((N-r)\theta)\sin(j\theta)\sin((j+N)\theta)\cr
\lambda&=2\cos(2\theta)\cr
}}
satisfies \lintau, the boundary conditions and the positivity requirement
$a^r_j>0$. Notice that \soltau\ respects the rank-level duality of
\lintau\ (exchange of $r\leftrightarrow j$ and $N\leftrightarrow f$).

We finally find $\tau=2N/(f+N)$. We then need to treat separately
the cases $\tau<1$, $\tau=1$, $\tau>1$, which leads to the following
discussion:

\noindent $\star$ For $f<N$, $1<\tau<2$ and the dominant correction is
the $e^{-\zeta}$ correction; as the next correction coming
from $\chi^r_f$ is of order $e^{-2\zeta}$, one can write
\eqn\asyFloToz{F^n_l\buildrel T\rightarrow 0\over = 
-T\left[ \log\chi^n_l(G_f)
+\gamma^n_l {T\over T_0} + a^n_l
\left( {T\over T_0}\right)^\tau +\ldots
\right]}
where $\gamma^n_l$ and $a$ (the constant in $a^n_l$) 
are functions of the ratio $B/T$, so that
\eqn\asyFloTBoz{\sigma^n_l\buildrel T\rightarrow 0\over\propto 
{1\over T_0} \left[ 1- cst\ \left( {T\over T_0}\right)^{N-f\over N+f}
\right]}
The susceptibility tends to a finite value, as in the screened case,
but the next correction has a non-integer power.

\noindent $\star$ For $f=N$ (the ``self-dual'', i.e. square diagram),
$\tau=1$:
the exponential corrections of the $\chi^r_f$ are exactly of the same
order as those of the $\chi^r_j$, $j<f$, so that they have to be inserted
as boundary conditions in \lintau: $a^r_f$ now has a non-zero
value which one can extract from \Fdilog\ (this new boundary condition 
breaks the self-duality of the equations). In fact,
there is no solution to this system of inhomogeneous linear equations,
precisely because $\lambda=2\cos(\pi/N)$ (i.e. $\tau=1$) is an eigenvalue
of the matrix of this system.
This is a ``resonance'' phenomenon, 
and as is usual in such degenerate cases
we must replace the ansatz \anstau\ with
\eqn\anstaub{\chi^r_j(\zeta)=\chi^r_j(+\infty) 
\left( 1+ (a^r_j+b^r_j\zeta)e^{-\zeta}\right)}
One can then find a solution to the corresponding inhomogeneous linear
system with $b^r_j$ of the type of \soltau: 
$b^r_j=b\, \sin(\pi r/N) \sin(\pi j/N)$
so that
\eqn\asyFloToy{F^n_l\buildrel T\rightarrow 0\over =
-T\left[ \log\chi^n_l(G_f)
+b^n_l {T\over T_0} \log\left( {T\over T_0}\right) + \ldots
\right]}
where again $b$ is a $B/T$-dependent constant;
\eqn\asyFloTBoy{\sigma^n_l\buildrel T\rightarrow 0\over\propto 
{1\over T_0} \log\left( {T\over T_0}\right)}
The susceptibility has a logarithmic divergence.

\noindent $\star$ For $f>N$, $\tau<1$, and we find:
\eqn\asyFloTox{F^n_l\buildrel T\rightarrow 0\over = 
-T\left[ \log\chi^n_l(G_f)
 +a^n_l \left({T\over T_0}\right)^\tau + \ldots \right]
}

\eqn\asyFloTBox{\sigma^n_l
\buildrel T\rightarrow 0\over\propto
{1\over T_0}
\left( {T\over T_0} \right)^{-{f-N\over N+f}}}
The susceptibility is divergent as $T\rightarrow 0$, but with
a non-integer power-law.

In all three cases, the fact that the divergence of the susceptibility is
always slower than $1/T$ indicates that the $SU(N)$ spin of the impurity
is completely screened at the IR fixed point (cf section 2).

\newsec{Interpretation of the entropy at zero temperature.}
We shall first discuss in an abstract setting, how to calculate
the entropic factor connected to a particle in a general theory.
We shall then apply this to the impurity in our model, which will yield
the zero temperature entropy (since the elctrons do not contribute to
it).

\subsec{General principles.}
We consider a system in which states can be characterized by a
quantum number $R$. As several states may have the same quantum number,
we associate to each $R$ a degeneracy $d_R$. 
Let us now imagine that we take a state
which has the quantum number $R$, and add to it a particle: we obtain a new
state which has the quantum number $R'$. Of course, for a given $R$ and
a given type of particle, not all $R'$ are allowed; therefore we associate to
the particle the adjacency
matrix $A_{RR'}$ which is $0$ if the transition from $R$ to $R'$ is forbidden,
and $1$ (or in fact any positive integer in case of multiple possible
transitions) if the transition is allowed.

Let us now start from the vacuum, which conventionally has the
quantum number $R=\emptyset$. If we put one particle with associated
adjacency matrix $(A_0)_{RR'}$ into the system, the number of states allowed is
by definition
\eqn\entfact{\Omega=\sum_{R} (A_0)_{\emptyset R} d_R}
This expression is correct for a system of fixed finite length $L$;
the entropy at zero temperature is equal to $\log\Omega$, that is
the logarithm of an integer.
However, we shall argue that this is in fact not the proper definition
of the entropic factor associated to this particle in the thermodynamic
limit. Indeed, as the size $L$ of the system grows, at fixed temperature
$T$, the average number of particles in the system necessarily diverges.
In fact in a massless theory, it is obvious that the number of
particles is of order $TL\gg 1$ (even if $T$ is in the end sent to $0$).

Therefore, we come to the conclusion that to define the entropic factor
of a given particle as $L\rightarrow\infty$, one should always consider
it as surrounded with a large number of other particles. One should
then extract from the entropy of the resulting state the contribution of
the particle we are interested in.

From this point of view the entropic factor is
\eqn\entfactb{\Omega=\sum_R (A_0 A_1^{N_1} A_2^{N_2} \ldots A_p^{N_p})
_{\emptyset R} d_R}
where we have introduced the numbers $N_i$ and adjacency matrices $A_i$
of the different types of particles.

We shall make the assumption that all the matrices $A_i$
commute. This will always be the case for us.
Then in the limit $N_i\rightarrow\infty$, we can write
\eqn\entfactc{\Omega\propto \lambda_0 \lambda_1^{N_1} \lambda_2^{N_2}
\ldots \lambda_p^{N_p}}
where $\lambda_i$ is the biggest eigenvalue of $A_i$, which is necessarily
real positive since $A_i$ has positive entries, and which is
associated with the Perron-Frobenius eigenvector common to all
matrices $A_i$.
The contribution to the entropy of our original particle has now become
$S=\log\lambda_0$.

An important remark is that
the entropy does not depend on the degeneracy numbers $d_R$~!
This may sound slightly paradoxical, but it will in fact play
a key role in what follows.

Let us come back to our Kondo model. Here we have in fact two quantum
numbers.
We shall first introduce and study
these two quantum numbers one at a time, constructing the adjacency matrices
and computing their Perron-Frobenius eigenvalues according to the procedure
described above.

\subsec{The $SU(N)$ quantum number.}
Let us suppose first that the quantum number
$R$ is simply the $SU(N)$ irreducible representation of the state. 
To a particle which
belongs to the representation $R_0$ we associate the 
adjacency matrix $(A_{R_0})_{RR'}$ which is the usual tensor product matrix in the
space of representations;
it is defined by the decomposition rule
$$R_0\otimes R = \bigoplus_{R'} (A_{R_0})_{RR'} R'$$
i.e. matrix elements of $A_{R_0}$
are $SU(N)$ Littlewood-Richardson coefficients.
The degeneracy $d_R$ is the dimension of the representation $R$.
Equation \entfact\ reads here
\eqn\entfactu{\Omega=\sum_R (A_{R_0})_{\emptyset R} d_R=d_{R_0}}
so that we immediately find the correct result that $S=\log\Omega$
is the logarithm of the dimension of the representation $R_0$.
Here, the formalism we developed is useless; it just gives in a complicated
way the same result since the analogue of Eq. \entfactb\ is
\eqn\entfactbu{\Omega=\sum_R (A_{R_0} A_{R_1}^{N_1} A_{R_2}^{N_2}\ldots
A_{R_p}^{N_p})_{\emptyset R}d_R=d_{R_0} d_{R_1}^{N_1} d_{R_2}^{N_2}
\ldots d_{R_p}^{N_p}}
which immediately exhibits the Perron-Frobenius eigenvalues
$\lambda_i=d_{R_i}$ of the adjacency matrices.

\subsec{The $SU_q(N)$ quantum number (restricted to level $f$).}
Let us first introduce the quantum group $U_q(\goth{sl}(N))$,
with $q=\exp(2i\theta)=\exp(2i\pi/(f+N))$. We remind the reader
that given a simple Lie algebra $\goth g$, one can construct
a family of deformations $U_q(\goth g)$ of it:
$U_q(\goth g)$ is an algebra (in fact a Hopf algebra,
see~\ref\FUC{For a physical
introduction to quantum groups, see\hfil\break
J.~Fuchs, {\it Quantum groups and affine Lie algebras}, Cambridge
University Press (1992).}) whose defining relations are deformations of
those of $\goth g$, with $q$ the deformation
parameter (when $q\rightarrow 1$ one recovers the undeformed
universal enveloping algebra $U(\goth g)$). Quantum groups are the natural
symmetries of integrable models, even though the $U_q(\goth{sl}(N)$
symmetry that we now consider is in a sense ``hidden'' in the Kondo
model, and we rediscover it by computing thermodynamic quantities.

Let us turn to the representation theory of $U_q(\goth{sl}(N))$.
For generic $q$, it is
identical to that of $\goth{sl}(N)$ (and irreducible representations
can be depicted e.g. by Young
tableaux as in appendix A). However, for $q$ a root of unity\foot{Here,
we consider the so-called ``restricted specialization'' of
$U_q(\goth{sl}(N))$},
the situation is more complicated: a subclass of irreducible
representations, the
``good'' representations behave like for any value of $q$; but
the others have a more complicated behavior (in particular several
Young tableaux merge together into indecomposable but
not irreducible representations). 
In our case, when
$q$ is a primitive $(f+N)^{\rm th}$ root of unity, the ``good''
representations, which we also call
``$SU_q(N)$ representations'' for simplicity,
are characterized by
the property that the number of columns of their Young tableau
$n^1+\cdots+n^{N-1}=m^1-m^N \le f$. We denote their set
by $P^+(N,f)$\foot{We borrow this notation
from affine Lie algebra theory, since $P^+(N,f)$ can also be defined
as the set of integral dominant weights of $A^{(1)}_{N-1}$ at level
$f$.}.

Fortunately, it can be shown that one can consistently
restrict the representation theory $U_q(\goth{sl}(N))$ to the
``good'' representations. This is the situation we consider now:
the $SU_q(N)$ quantum number is precisely an element of the set $P^+(N,f)$.
The justification of this choice is that, from the form of the B.A.E.,
we conjecture that in the spin sector of the Kondo model,
only such representations are present. In fact, this can also be
guessed from the similar situation in the WZW $SU(N)$ level $f$ theory.

Let us comment on the physical implication of such a quantum
number. One-particle states are now organized in multiplets of
$SU_q(N)$ with certain ``good'' representations,
(for us, they are the fundamental representations of $SU_q(N)$);
however, as soon
as one considers multi-particle states, it becomes clear that the Hilbert
space of our theory does not have a Fock structure any more: indeed
by performing the tensor product of
enough representations of $P^+(N,f)$ one necessarily obtains
representations with more than $f$ columns and which therefore do no belong
to $P^+(N,f)$. A truncation is needed to keep only the ``good''
representations in the tensor product.

The non-Fock structure of the states
is most easily imagined by considering the physical particles as
kinks~\ref\F{P.~Fendley, {\it Phys. Rev. Lett.} 71 (1993), 2485.},
that is states
interpolating between different classical vacua labeled by the
$SU_q(N)$ quantum number.
This is the same ``kink structure''
which appears, for example, in RSOS models~\ref\KM{For similar kink
structures, see for example\hfil\break
T.~Klassen and E.~Melzer, {\it Nucl. Phys.} B382 (1992), 441\semi
P.~Fendley, H.~Saleur, Al.B.~Zamolodchikov,
{\it Int. J. Mod. Phys.} A8 (1993), 5751.}. In the end, this non-Fock
structure will be responsible for the fractional zero temperature
entropy.

This truncation (and the associated kink structure)
is apparent when one builds the adjacency matrices
$(A^r_j)_{RR'}$; they are the {\it truncated tensor product matrices}
for the quantum group $U_q(\goth{sl}(N))$ associated to the rectangular
Young tableaux $r\times j$. In the kink picture, the adjacency
matrices describe the
possible transitions between the different classical vacua.
Here we do not give an abstract algebraic
construction for $A^r_j$ but rather a simple explicit procedure.

We start
with $A^1_1$ which is obtained by taking the corresponding
standard (i.e. untruncated) tensor product matrix and restricting it 
(in the naive sense) to $P_+(N,f)$.
Explicitly, this means that $(A^1_1)_{RR'}=1$
if there is a $a\in \{ 1,\ldots, N\}$ such that $R'=R+e^a$, $0$ otherwise.
We have shown in fig.~\gra\ the set $P_+(3,3)$; each arrow corresponds
to the addition of a $e^a$, $a=1\ldots 3$, so that $A^1_1$ is precisely
the adjacency matrix of the resulting graph.
\fig\gra{The set $P_+(3,3)$ and the graph
of $A^1_1$.}{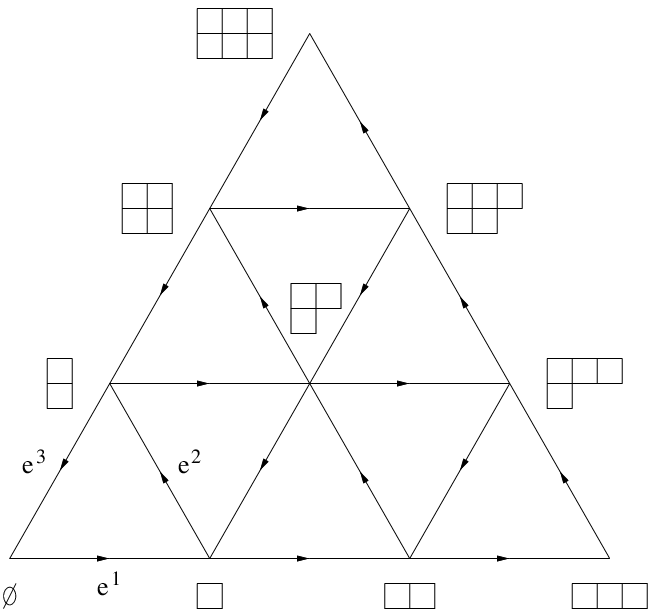} 

This way we have defined the matrix $A^1_1$. We could use the same
restriction procedure
for any matrix $A^r_1$ corresponding to a fundamental representation
of $SU(N)$; explicitly, $(A^r_1)_{RR'}=1$ if there are
$a_1$, $\ldots$, $a_r$ all distinct such that $R'=R+e^{a_1}+\cdots
+e^{a_r}$, $0$ otherwise. Fig.~\grab\ gives an example of such a construction.
\fig\grab{The set $P_+(4,2)$ and the graph of $A^2_1$, which consists of
the edges with double arrows.}{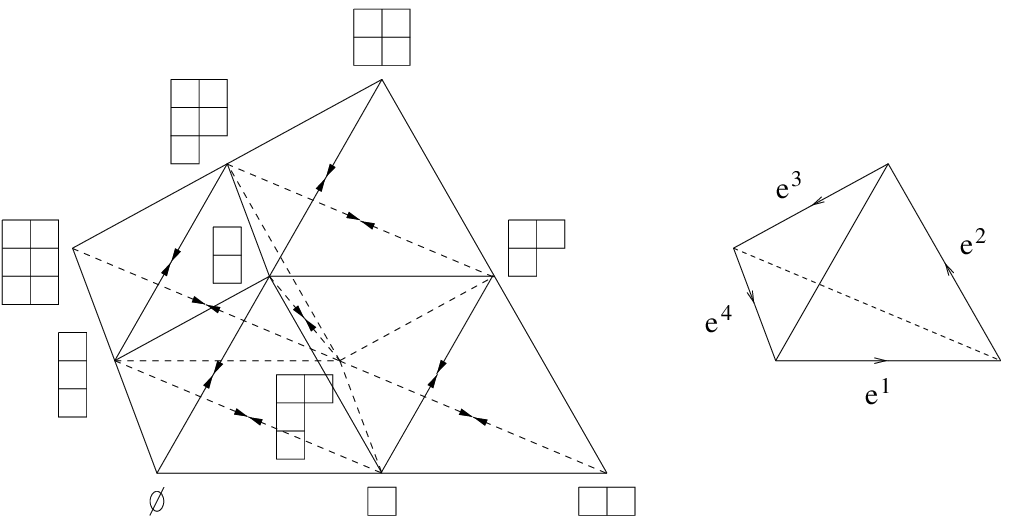} 

The higher matrices $A^r_j$, $j>1$
cannot be obtained in the same way; one must use the fusion
procedure. In the case of rectangular Young tableaux,
the matrices $A^r_j$ can be defined by imposing the by now familiar
fusion equations
\eqn\fusA{A^r_j A^r_j = A^{r+1}_j A^{r-1}_j + A^r_{j+1} A^r_{j-1}}
with the boundary conditions $A^0_j=A^N_j=A^r_0=1$. It should be noted
that the fusion procedure requires only $A^1_1$ to build all the other
$A^r_j$; however, the sole use of eqs. \fusA\ requires all the
$A^r_1$, so as to have proper initial conditions to the recursion
$A^r_j=f(A^q_{j-1},A^{q'}_{j-2})$. In
fact one has the following explicit solution in determinant form:
\eqn\fusAb{A^r_j=\det(A^{r-a+b}_1)_{1\le a,b\le j}}
(where $A^r_1\equiv 0$ for $r<0$ or $r>N$).

One can finally check that the largest eigenvalue of $A^r_j$,
which like all eigenvalues is a $SU(N)$ character, is precisely
$\chi^r_j(G_f)$\foot{In the quantum group terminology this number
is called the {\it quantum dimension} of the representation $r\times j$.}.

\subsec{The Kondo model.}
In the Kondo model, as already mentioned,
the two quantum numbers described above naturally
appear in the Bethe Ansatz description through the appearance of
the $j$-strings for $1\le j<f$ and $j>f$. We shall now try
to apply the general method outlined above to the
impurity {\it in interaction} with the electrons, i.e. the (under,
over)screened impurity.
By inspection of the ground state (section 2), we can infer that
the impurity never has both adjacency matrices non-trivial: for the
underscreened case only the unrestricted $SU(N)$ adjacency matrix is non-trivial,
while the $SU_q(N)$ adjacency matrix is the identity matrix, and
the converse statement for the overscreened case. This means that we
can hide one of the two quantum numbers inside the degeneracies $d_R$ of
the other; as we have already noted that these numbers $d_R$ play no role
in the thermodynamic limit, we can blindly apply the results of the two
previous subsections:

\blob Underscreened case ($f<l$). We apply the result of section 5.2
to the impurity, which we assume belongs to the representation
$n\times (l-f)$ due to the screening of the electrons.
Our procedure gives the correct result that the entropy at zero temperature
is the logarithm of the dimension of this representation.

\blob Overscreened case ($f>l$). We assume that the adjacency matrix
corresponding to the impurity is the $SU_q(N)$ truncated tensor product matrix
$A^{N-n}_l$; this non-trivial
adjacency matrix is due to the $(N-n)\times l$ electrons (or more precisely their
spin sector component) on the site of the
impurity, which are screening it: indeed to completely screen the spin of the impurity
(as was found in section 2) one needs precisely $(N-n)\times l$ electrons.

In this case too, one finds the correct
result that the entropy is the logarithm 
of $\chi^{N-n}_l(G_f)=\chi^n_l(G_f)$.

\bigbreak\bigskip\bigskip\centerline{{\bf Acknowledgements}}\nobreak
The authors are grateful to A.~Jerez for valuable
comments and discussions.
P.~Z-J. wishes to thank V.~Kazakov, H.~De~Vega, D.~Bernard for
discussions.

\appendix{A}{$SU(N)$ representations and characters}
An irreducible representation (irrep)
$R$ of $U(N)$ is characterized by its highest
weights, which form a sequence of $N$ integers $m^1\ge m^2
\ge \ldots\ge m^N$. $m^a$ is the length of the $a^{\rm th}$ row of the
corresponding Young tableau (fig.~\genY).
\fig\genY{A generic Young tableau.}{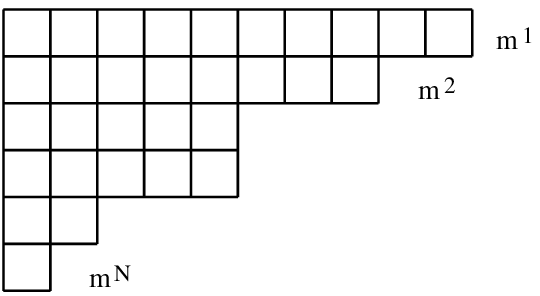} 
We shall write:
\eqn\yd{R=\sum_{a=1}^N m^a e^a}
As we are dealing with $SU(N)$ (and not $U(N)$)
irreps, we simply add the rule
that two irreps whose highest weights only differ by a constant are
equivalent. This amounts to imposing the relation:
$e^1+e^2+\cdots+e^N=0$.
One possible convention is then to suppose that $m^N=0$,
the other highest weights $m^a$ being non-negative (as on 
figs.~\gra\ and \grab).

One can also introduce another set of numbers to describe $R$: 
one defines $n^r$ to be the number of columns of given size $r$, $1\le r\le
N-1$. This corresponds to the decomposition
\eqn\ydb{R=\sum_{r=1}^{N-1} n^r \omega^r}
where $\omega^r=\sum_{a=1}^r e_a$ is the $r^{\rm th}$ fundamental weight
of $SU(N)$.
We have $n^r=m^r-m^{r+1}$.
In the case of a Bethe Ansatz state,
using Eq. \rep, one can show that
\eqn\NMrel{n^r=n^r_0-2\sum_q C^{qr} M^q}
where $n^r_0$ is the contribution of $R_0$,
and
$2C^{qr}=2C^{qr}(\kappa=0)=2\delta^{qr}-(\delta^{qr+1}+\delta^{qr-1})$
is the usual Cartan matrix of $A_{N-1}$. Rewritten explicitly,
this gives Eq. \qn.

To a representation $R$ and an element $G$ of $SU(N)$ (or, by analytic
continuation, of $SL(N)$) we
associate the character $\chi_R(G)$
defined as the trace of $G$ in the representation $R$. It
satisfies the basic $SU(N)$ invariance property:
$$\chi_R(G)=\chi_R(\Omega G\Omega^{-1})\quad\forall\Omega\in SU(N)$$
which implies that it only depends on the eigenvalues $z^a$ ($1\le a\le N$)
of $G$. For irreps, one has the following explicit formula:
\eqn\chiexpl{\chi_R(G)={\det((z^a)^{m^b+N-b})_{a,b=1\ldots N}
\over\det((z^a)^{N-b})_{a,b=1\ldots N}}}
where $\det((z^a)^{N-b})\equiv\Delta$
is simply the Van der Monde determinant of the $z^a$.

Let us now suppose that the $z^a$ are real positive, so that
one can order them: $z^1>z^2>\ldots>z^N$. Let us furthermore assume
that the representation $R$ becomes large, in the sense that
the $n^r\gg 1$. Then one can replace \chiexpl\ with the following
estimate:
\eqn\chiasya{\chi_R(G)\sim {1\over\Delta} 
\prod_{a=1}^N (z^a)^{m^a+N-a}}
Setting $z^a=\exp(-B^a/T)$ and expressing the $m^a$ in terms of the
$M^r$, one obtains Eq. \chiasy. 

For a large rectangular Young tableau $r\times j$, we have
$n^q=j\delta^{qr}$ so that only $n^r\gg 1$. A more careful estimate
gives
\eqn\chiasyb{\chi^r_j(G)
\buildrel j\rightarrow\infty\over\sim
{\Delta_1\Delta_2\over\Delta}
\prod_{a=1}^N (z^a)^{m^a+N-a}}
where $\Delta_1$ (resp. $\Delta_2$)
is the Van der Monde of the $z^a$ for $1\le a\le r$
(resp. $r+1\le a\le N$). It is quite close to \chiasya\ 
in the sense that the character is still approximated
by a constant times the
highest weight vector contribution.

More explicitly, for $z^a=\exp(-B^a/T)$, we have
\eqn\chiasyc{\chi^r_j(e^{-B/T}) 
\buildrel j\rightarrow\infty \over \propto
e^{-j\sum_{a=1}^r B^a/T}} 
(compare with Eq. \asyjb).

Let us finally expand $\chi_R(e^{-B/T})$ for $B\ll T$, i.e. around
$\chi_R(1_{SU(N)})$. Due to the $SU(N)$-invariance, the linear term
is necessarily proportional to the only linear invariant $\sum_{a=1}^N B^a$,
which in our case is zero. In the same way, there are only two quadratic
invariants, $(\sum_{a=1}^N B^a)^2=0$ and $\sum_{a=1}^N (B^a)^2$, so that
\eqn\chiloB{\log\chi_R(e^{-B/T}) = \log\chi_R(1_{SU(N)}) + {C\over 2}
\sum_{a=1}^N (B^a/T)^2+\ldots}
This motivates the definition of the magnetic susceptibility given
in \defsusc.
To compute the constant $C$, one uses the well-known identity:
$\Delta\chi_r=C_2(R)\chi_R$ where $\Delta$ is the Laplacian on $SU(N)$
and $C_2(R)$ is the quadratic Casimir of the irrep $R$. With the
usual normalizations this means that $C=C_2(R)/(N^2-1)$.

\appendix{B}{Computation of the free energy of the electrons}
This appendix presents the computation (from the Bethe Ansatz),
at given temperature $T$ and
magnetic field $B$, of the free energy per unit length of the electrons
\eqn\BFe{{\cal F}=-D T \sum_r \int d\Lambda\, G^{r1}
\star s(\Lambda-1) \log(1+\eta^r_f(\Lambda))}
We shall take the scaling limit only at the end of the calculations.

The key observation is that $2\pi D\delta_{j,f}G^{r1}\star s(\Lambda-1)
=(d/d\Lambda)\tilde{g}^r_j$ (cf \enb, \enbb); replacing $\tilde{g}^r_j$
with the l.h.s. of \TBAb, and inserting this into \BFe\ gives
\eqn\BFeb{{\cal F}=-{T^2\over 2\pi}\sum_{r,j}\int d\Lambda
{d\over d\Lambda}\left(\log(1+(\eta^r_j)^{-1})
-\sum_{q,k} G^{qr}\star C_{jk}\star \log(1+\eta^q_k) \right)
\log(1+\eta^r_j(\Lambda))}
The double integral which appears in \BFeb
\eqn\dubint{I\equiv\sum_{j,k,q,r}\int d\Lambda \, d\Lambda'\,
A^{qr}_{jk}(\Lambda-\Lambda') \log(1+\eta^r_j(\Lambda))
\log(1+\eta^q_k(\Lambda'))}
where $A^{qr}_{jk}\equiv (d/d\Lambda)(G^{qr}\star C_{jk})$, would seem
to vanish due to the change of sign under the simultaneous
transformations $q\leftrightarrow r$, $j\leftrightarrow k$, $\Lambda
\leftrightarrow -\Lambda$. In fact it does not, because of
the non-uniform convergence of this integral as $\Lambda$, 
$\Lambda'\rightarrow\pm\infty$, and the correct result is:
\eqn\dubintb{
I={1\over 2}\sum_{j,k,q,r}G^{qr}(\kappa=0) C_{jk}(\kappa=0)
\bigg[ \log(1+\eta^r_j)\log(1+\eta^q_k) \bigg] ^{+\infty}_{-\infty}
}
We now use Eq. \TBA\ in the limit $\Lambda\rightarrow\pm\infty$
\eqn\TBAlim{\log(1+\eta^r_j(\pm\infty))-\sum_{q,k} C^{qr}(\kappa=0)
G_{jk}(\kappa=0)
\log(1+(\eta^q_k(\pm\infty))^{-1})={g^r_j(\pm\infty)\over T}}
to express in a more suggestive form $I$:
\eqn\dubintc{\eqalign{
I&={1\over 2} \sum_{j,r}
\bigg[\log(1+\eta^r_j)\log(1+(\eta^r_j)^{-1}) \bigg]^{+\infty}_{-\infty}\cr
&+{1\over 2} \sum_{j,k,q,r} G^{qr}(\kappa=0) C_{jk}(\kappa=0)
\bigg[ {g^r_j\over T} \log(1+\eta^q_k) \bigg]^{+\infty}_{-\infty}\cr
}}
The free energy then takes the form
\eqn\BFec{\eqalign{
{\cal F}= &-{T^2\over 2\pi} \sum_{j,r}
\left[ -{1\over 2}\int \left[ d( \log(1+(\eta^r_j)^{-1}) ) \log(1+\eta^r_j)
+ d( \log(1+\eta^r_j) ) \log(1+(\eta^r_j)^{-1}) \right] \right] \cr
&+{T\over 4\pi}\sum_{j,k,q,r} G^{qr} C_{jk}
\bigg[ g^r_j \log(1+\eta^q_k) \bigg]^{+\infty}_{-\infty}\cr
}}
($\kappa=0$ is implied for all kernels now). We shall compute this expression
in the scaling limit.

The first term is usually written in terms of
Rogers' dilogarithmic function $L$, by taking $1/(1+\eta)$ as new variable;
in fact one can easily show that, due to cancellations
between $\Lambda=+\infty$ and $\Lambda=-\infty$,
the quantity between the brackets is precisely
equal to the following (known) dilogarithm sum~\ref\Ki{
A.N.~Kirillov, {\it Zap. Nauch. Semin. LOMI} 164 (1987), 121.}:
\eqn\dilog{
\sum_{j=1}^f \sum_{r=1}^{N-1} 
L \left( {1\over 1+\eta^r_j(+\infty)} \right)
= {\pi^2\over 6} {f\over N+f} (N^2-1)
}

Concerning the second term, 
one might be tempted to replace $\sum_{r,j} G^{qr} C_{jk} \,
g^r_j(\pm\infty)$ with its value $\tilde{g}^q_k(\pm\infty)$
and then evaluate the sum,
which turns out to be zero. This is in fact incorrect because of
the divergences which occur at $j=\infty$ due to the presence of the
magnetic field $B$. The correct procedure is to isolate the term
where $B$ appears
\eqn\BFeB{{\cal F}_B \equiv {T\over 4\pi}
\sum_{q,r} G^{qr} b^r \sum_{k=1}^\infty
\bigg[ \log(1+\eta^q_k) \bigg]^{+\infty}_{-\infty}
\sum_{j=1}^J j \, C_{jk}}
where we have introduced a cutoff $J$ for the last sum (which would naively
vanish as $J\rightarrow\infty$). Using $\sum_{j=1}^J j\, C_{jk}=
\delta_{k J} (J+1)/2 - \delta_{k J+1} J/2$
and \defch\ we find:
\eqn\BFeBb{{\cal F}_B = {T\over 4\pi} \lim_{J\rightarrow\infty}
\sum_r b^r \bigg[ (J+1) \log\chi^r_J - J \log\chi^r_{J+1} \bigg]
^{+\infty}_{-\infty}
}
The explicit expressions $\chi^r_j(-\infty)=\chi^r_j(e^{-B/T})$,
$\chi^r_j(+\infty)=\chi^r_{j-f}(e^{-B/T})$ ($j>f$), and the estimate
\chiasyc\ allow to compute \BFeBb; the result is:
\eqn\BFeBc{{\cal F}_B=-{1\over 4\pi} f \sum_{a=1}^N (B^a)^2}

Putting \dilog\ and \BFeBc\ together, we obtain the final expression
for the free energy:
\eqn\Bfinal{{\cal F}=-T^2 {\pi\over 12} {f\over N+f} (N^2-1)
-{1\over 4\pi} f \sum_{a=1}^N (B^a)^2}

Let us comment on this result. The first term is the central charge
term for a theory with only one chirality and central charge
$c=(N^2-1)f/(N+f)$: this is what we expect since the spin sector
of the electrons is a WZW $SU(N)$ level $f$ theory. Of course once
one puts together the spin, flavor and charge sectors, one recovers
the total central charge
\eqn\cc{c=(N^2-1){f\over N+f} + (f^2-1){N\over N+f} + 1 = Nf}
for $N\,f$ free complex fermions.

The second term can be directly obtained as the free energy of
the electrons in the magnetic field $B$: indeed the flavor and
charge sectors do not contribute to it. One has the simple
formula
\eqn\Bmag{{\cal F_B}=-T f\int_{-{\cal D}}^{+\infty} {dk\over 2\pi}
\sum_{a=1}^N \log(1+e^{-(k+B^a)/T})}
where ${\cal D}$ is some momentum cutoff of the order of $D$.
The factor of $f$ comes from the $f$ flavors. After sending
${\cal D}$ to infinity and discarding the divergent $T$-independent
part, one obtains \BFeBc.

\listrefs

\bye